\documentclass[10pt]{article}
\pdfoutput=1

\usepackage{amsmath,amssymb,amsfonts,amscd,amssymb,mathrsfs}
\usepackage{xcolor}
\definecolor{darkblue}{rgb}{0.1,0.1,.7}
\usepackage[colorlinks, linkcolor=darkblue, citecolor=darkblue, urlcolor=darkblue, linktocpage,pagebackref]{hyperref} 
\usepackage[square, comma, compress,numbers]{natbib}
\usepackage[]{graphicx}
\usepackage{geometry}
\usepackage[margin=10pt,font=small,labelfont=bf]{caption}
\usepackage{ifthen}
\usepackage{tikz}
\usepackage{subcaption}
\usepackage{booktabs,multirow}
\usepackage{hhline}
\usepackage{tablefootnote}
\usepackage[T1]{fontenc}
\usepackage{bm}
\usepackage{tikz-cd}
\usepackage{soul}
\usepackage[utf8]{inputenc}
\usepackage{marginnote}
\usepackage{dsfont} 
\usepackage{titlesec}
\titleformat*{\section}{\large\bfseries}
\titleformat*{\subsection}{\normalsize\bfseries}
\titleformat*{\subsubsection}{\normalsize\it}
\titleformat*{\paragraph}{\normalsize\bfseries}
\titleformat*{\subparagraph}{\normalsize\bfseries}

\usepackage{setspace}
\usepackage{amsmath, amssymb, amsthm, float, graphicx}

\usepackage{mymacros}

\usepackage{ytableau}
\usepackage[vcentermath]{youngtab}

\numberwithin{equation}{section}

\usepackage{mymacros}

\begin{document}
	
	\vspace*{-.6in} \thispagestyle{empty}
	\begin{flushright}
	\end{flushright}
	\vspace{.2in} {\Large
		\begin{center}
			\resizebox{\textwidth}{!}{\bf  Random Field Ising Model and Parisi-Sourlas Supersymmetry}\\
						{\bf I. Supersymmetric  CFT}\\
		\end{center}
	}
	\vspace{.2in}
	\begin{center}
		{\bf 
			Apratim Kaviraj$^{a,b}$, \ Slava Rychkov$^{c,b}$,  \ Emilio Trevisani$^{b}$
		} 
		\\
		\vspace{.2in} 
		{\it $^{a}$Institut de Physique Th\'{e}orique Philippe Meyer, \\ $^{b}$
			Laboratoire de Physique de l’Ecole normale sup\'erieure, ENS, \\
			Universit\'e PSL, CNRS, Sorbonne Universit\'e, Universit\'e de Paris, F-75005 Paris, France}\\
		{\it $^{c}$ Institut des Hautes \'Etudes Scientifiques, Bures-sur-Yvette, France}\\
	\end{center}
	
	\vspace{.2in}
	
	\begin{abstract}
		Quenched disorder is very important but notoriously hard. In 1979, Parisi and Sourlas proposed an interesting and powerful conjecture about the infrared fixed points with random field type of disorder: such fixed points should possess an unusual supersymmetry, by which they reduce in two less spatial dimensions to usual non-supersymmetric non-disordered fixed points. This conjecture however is known to fail in some simple cases, but there is no consensus on why this happens. 
		In this paper we give new non-perturbative arguments for dimensional reduction.
		We recast the problem in the language of Conformal Field Theory (CFT). We then exhibit a map of operators and correlation functions from Parisi-Sourlas supersymmetric CFT in $d$ dimensions to a $(d-2)$-dimensional ordinary CFT. The reduced theory is local, i.e.~it has a local conserved stress tensor operator. As required by reduction, we show a perfect match between superconformal blocks and the usual conformal blocks in two dimensions lower. This also leads to a new relation between conformal blocks across dimensions. This paper concerns the second half of the Parisi-Sourlas conjecture, while the first half (existence of a supersymmetric fixed point) will be examined in a companion work.
		\end{abstract}
\vspace{.2in}
\vspace{.3in}
\hspace{0.7cm} December 2019
	\clearpage
	
	\tableofcontents

\section{Introduction }\label{sec:intro}

Physical systems realized in nature often have some kind of random impurities.
The presence of such impurities may change the behavior of a system.
It is therefore of great importance to understand how these changes can occur.
In order to investigate this question one typically considers a statistical model and adds a disorder  interaction. Physical observables are then computed by averaging over the disorder. 

Depending on the physical system there are two types of averaging possible. If the impurities achieve thermal equilibrium, they have to be treated like another degree of freedom.  
Hence a sum over disorder configurations should be included in the partition function.
This is called \textit{annealed} disorder.

In this paper we are concerned with the second case, that of \textit{quenched} disorder. Here impurities are not in thermal equilibrium. The observables (e.g. correlation functions) are computed for a fixed disorder configuration and the disorder average is performed in the end.

We will consider a specific class of quenched disordered theories, where a disorder field is coupled to a local order parameter. An interesting example of this, which has a wide range of physical applications, is the Random Field Ising Model (RFIM). One can think of this as the regular Ising model in a random magnetic field (or in presence of random magnetic impurities). This has the Hamiltonian 
\be
\label{Ham}
\mathcal{H}=-J \sum_{\langle ij\rangle} s_i s_j +h_i s_i\,,
\ee
with $s_i = \pm 1$. The $h_i$ is a random magnetic field. It is drawn from some distribution, which we choose to be Gaussian with zero mean $\overline{h_i} = 0$, and is characterized by the variance  $\overline{h_i h_j}  = H \d_{ij}$.\footnote{Averages over $h$ will be denoted by an overline.} For a nice review on this topic see chapter 8 of \cite{cardy_1996}.
\begin{figure}[t]
	\begin{center}
		\includegraphics[width=0.6\textwidth]{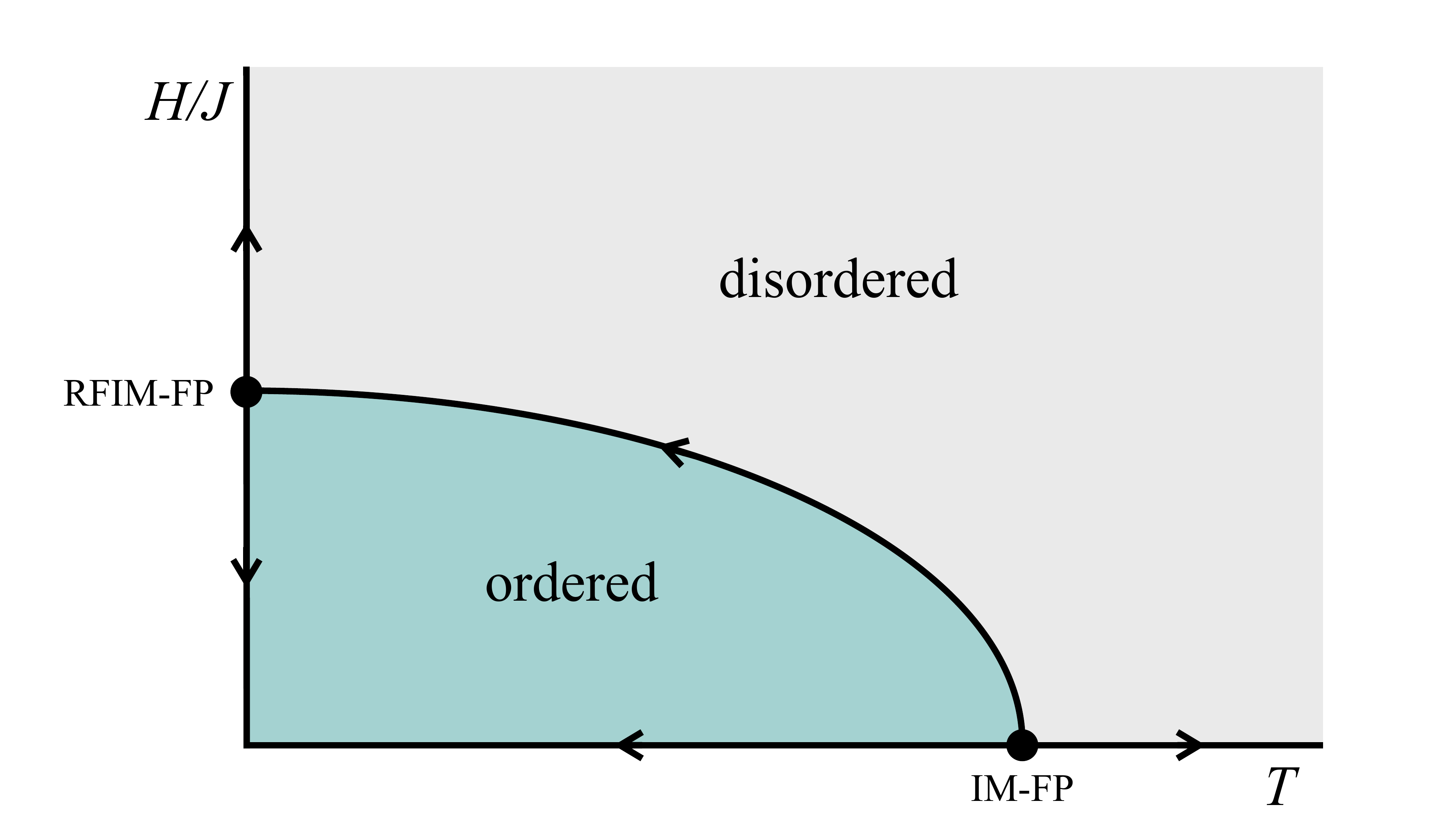}
	\end{center}
	\caption{
		The RFIM phase diagram for $d>2$. 
		\label{rfimphase}}
\end{figure}

For a fixed configuration of $h_i$, one studies the thermal fluctuations of the spins at a finite temperature, and then averages over $\{h_i\}$. The presence of random field drastically changes critical behavior, in particular the lower and upper critical dimensions are shifted to 2 and 6. 
For $d\leq 2$, the system is in the disordered phase ($\overline{\, |\langle s \rangle|\, }=0$) at any temperature.
This is dictated by the Imry-Ma criterion \cite{PhysRevLett.35.1399}, from an estimate of the free energy cost due to the flipping of a domain wall. For $d>2$, the phase diagram is as in Figure \ref{rfimphase}. There is an ordered phase and a disordered phase separated by a second-order phase transition.
From the two fixed points shown in the figure, the IM-FP controls the usual critical behavior (no quenched disorder), while RFIM-FP controls the critical behavior with quenched disorder turned on.
The interesting case is $2<d<6$, when the latter fixed point is non-gaussian.

The continuous version of RFIM can be realized with a scalar theory having a $\phi^4$ interaction:
\be
\int \dd^d x \ [ \frac{1}{2}  ( \partial\phi(x) )^2  + m^2 \phi^2(x)+ \l \phi^4(x) + h(x) \phi(x) ] \, .  \label{RFIM} 
\ee
One can study its fixed point by perturbation theory, in $d=6-\e$ dimension. This yields the surprising result\footnote{This was first noticed by Aharony, Imry and Ma in 1976 \cite{Aharony:1976jx}. They gave a strong evidence for this connection by showing  that at any order in perturbation theory the most IR-divergent Feynman diagrams of the random field theory can be equivalently written in terms of diagrams of a $d-2$ dimensional theory.}
that the critical exponents are exactly equal to those of the usual Wilson-Fisher fixed point in $d=4-\e$. 

\begin{figure}[t]
	\graphicspath{{Fig/}}
	\def\svgwidth{0.6\textwidth}
	\centering
	\input{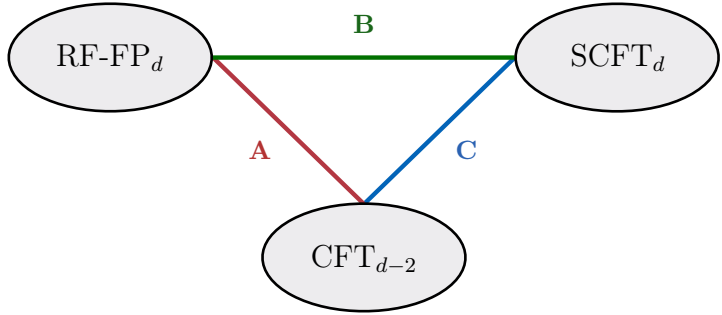} 
	\caption{
		The Parisi-Sourlas conjecture (our schematic formulation).
		\label{triangle}}
\end{figure}

An explanation of this is provided by a remarkable conjecture due to Parisi and Sourlas  in 1979 \cite{Parisi:1979ka}.\footnote{Parisi and Sourlas have stated the conjecture in different terms, in particular without mentioning CFTs. We propose a natural reformulation in modern language, which we will find useful.}
The idea is that there is a relation between the IR fixed points of three seemingly unrelated models:
\begin{itemize}
	\item The fixed point of a random field model in $d$ dimensions (RF-FP$_d$),
	\item The fixed point of a supersymmetric field theory without disorder in $d$ dimensions (SCFT$_d$),
	\item The fixed point of a model without disorder in $d-2$ dimensions (CFT$_{d-2}$),
\end{itemize}
as we show in Figure \ref{triangle}. 
The link $\aaa$  in Figure \ref{triangle} is intriguing, as
it relates the fixed point of a complicated disordered model to the fixed point of a pure system (which can be described by standard techniques). 
Even more surprisingly, the pure theory lives in two less dimensions, providing an unexpected relation between different universality classes across dimensions.
The Parisi and Sourlas conjecture thus explains the link $\aaa$ through $\bbb$ followed by $\ccc$.

Link $\bbb$ in Figure \ref{triangle} means that the fixed point of a random field theory should possess an enhanced symmetry, called Parisi-Sourlas supersymmetry. Theories with this kind of supersymmetry have very unusual features, at least to a high energy physicist. Most notably they violate spin-statistics, as their anticommuting degrees of freedom transform as scalars of the rotation group (i.e.~fermions have no spinorial indices). In particular, they are non-unitary. For the case of RFIM, the SUSY theory in $d$ dimensions is given by
\be
\int \dd^d x\dd\thetab \dd\theta \ [\frac{1}{2}  \partial^a\Phi(x,\theta,\thetab) \partial_a \Phi(x,\theta,\thetab)+m^2 \Phi^2(x,\theta,\thetab)+  \l \Phi^4(x,\theta,\thetab) ] \ ,
\label{PSIMd}
\ee
where $\Phi$ is a superfield and $\partial^a$ is the derivative in superspace. This will be discussed in detail in section \ref{PSrev}.

Finally, link $\ccc$ in Figure \ref{triangle} is called ``dimensional reduction''. 
It says that the correlation functions of the SUSY theory reduce to those of a $d-2$ dimensional theory, which is identical to the disordered model but without the disorder interaction.  The dimensional reduction heavily relies on the special supersymmetry of the theory. Roughly speaking in this supersymmetry two fermionic degrees of freedom eat two bosonic ones, effectively reducing the dimensionality of the system. This mechanism is reminiscent to what happens in Gaussian theories, where fermionic integrals cancel bosonic ones. As a final remark, we point out  that dimensional reduction was proposed not only for the IR fixed point, but also all along the RG flow, thus relating  quantum field theories in $d$ and $d-2$ dimensions.

In some examples, non-perturbative studies have verified some aspects of the Parisi-Sourlas conjecture. 
In particular, the critical point of branched polymers in $2<d<8$ has been shown to undergo dimensional reduction for all integer $d$ \cite{PhysRevLett.46.871, zbMATH02068689,CardyLecture}, confirming its equivalence to the Lee-Yang class of fixed points in $d-2$ dimensions. To be precise, these studies verified directly the $\aaa$ link of the conjecture by establishing an equivalence at the microscopic level, without considering the $\bbb$ and $\ccc$ links.

On the contrary, the conjecture was found to imply wrong results for the random field Ising model in $d=3$ and $4$. E.g.~according to the conjecture $\text{RFIM}_{d=3}$ should be related to Ising model at $d=1$, but this does not work: the latter does not even have a phase transition, but the former does, by the Imry-Ma criterion and by the rigorous results  \cite{ImbrieLCD, ImbrieLCD2}. In the $4\to 2$ case there is also a discrepancy: both models have a phase transition, but the critical exponents of RFIM in $d=4$ are found to be significantly far from critical Ising model in $d=2$ \cite{Picco1}. On the positive side, dimensional reduction and supersymmetry predictions for the case of $5\to 3$ have been confirmed by recent numerical studies \cite{Picco2, Picco3}.

There seems to be no consensus on how the conjecture works or fails, prompting us to undertake our own study.\footnote{There are several proposals in literature to explain its failure. State of the art will be reviewed in a companion paper \cite{paper2}. Our impression from studying the vast literature is that link $\ccc$ is usually considered more trustworthy than $\bbb$.}
Link $\bbb$ will be studied in the companion paper \cite{paper2}, where the RG flow of the random field theories is analyzed to see if it leads to supersymmetric fixed points. Here instead we will focus on $\ccc$, giving new non-perturbative arguments in its favor. Indeed, one gap in most justifications of $\ccc$ that we have seen is that they rely on a weakly  coupled Lagrangian description, while the fixed points in question become strongly coupled unless $d$ is close to 6. Our aim will be to close this gap.

Our main tool will be to reformulate the problem axiomatically, using the modern language of conformal field theory (CFT). This language is non-perturbative, so it lends itself naturally to our task. Using this language, 
we will illustrate the properties of conformal field theories with Parisi-Sourlas supersymmetry, and check the absence of any pathological feature. We will then explain the meaning of dimensional reduction in axiomatic CFT, and what this implies.

The paper is organized as follows.
We start by reviewing the Parisi-Sourlas conjecture in section \ref{PSrev}. In section \ref{sec:susy} we develop the non-perturbative theory of Parisi-Sourlas supersymmetric fixed points using the CFT language. In section \ref{sec:dimred}, we show how dimensional reduction takes place in Parisi-Sourlas CFTs. 
We show that the $d-2$ dimensional theory is local, and that the superconformal blocks of the Parisi-Sourlas CFT in $d$ dimensions are equal to standard conformal blocks in $d-2$ dimensions. As a byproduct, we find a nice formula relating conformal blocks in $d$ and $d-2$ dimensions. We end the paper with some concluding remarks in section \ref{sec:conclusions}.

\section{Review of the  Parisi-Sourlas conjecture }\label{PSrev}

\subsection{From random fields to Parisi-Sourlas supersymmetry }
In the introduction we explained that the random field models are conjectured to have an IR fixed point with enhanced supersymmetry \cite{Parisi:1979ka} (see also \cite{Parisi:1982ud, MCCLAIN1983430}). 
The aim of this section is to review how this conjecture comes about.
In the following we describe the main ingredients that we need for this---the replica method and a field redefinition introduced by Cardy.

Let us start with the theory of a single scalar field $\phi$ coupled to a random quenched magnetic field $h(x)$. The partition function of the model is defined by
\be\label{Sh}
Z_h=\int \mathcal{D}\phi e^{-\mathcal S[\phi,h]} \ ,
\qquad 
\mathcal  S[\phi,h]=\int \dd^dx \big[\frac{1}{2}(\partial \phi)^2 + V(\phi) + h(x) \phi \big]\,, 
\ee
where we are mostly interested in the case where  the potential is $V(\phi)=m^2\phi^2+\l \phi^4$ which defines the random field Ising model (another interesting case is the cubic potential which describes the random field Lee-Yang fixed point).\footnote{The mass term has to be finetuned to reach the fixed point. This physically corresponds to tuning the temperature to the transition. The finetuned valued of $m^2$ is regulator dependent and generally nonzero (although it would be zero in dimensional regularization). } 
The partition function $Z_h$ depends on the shape of $h(x)$, which is sampled from a distribution $P(h)$ with zero mean $\overline{h(x)}=0$ and no spacial correlation $\overline{h(x)h(x')}= \kk \d(x-x')$.
Observables like correlation functions are computed by averaging over the disorder:
\be
\label{formula:1copy}
\overline{\langle A(\phi)   \rangle }= \int \mathcal{D}h P(h) \ \frac{\int \mathcal{D}\phi \ A(\phi)  \ e^{-\mathcal  S[\phi,h]}}{Z_h} \ .
\ee
Here $\langle A(\phi)\rangle$ denotes any correlation function built out of $\phi$'s, e.g.~$\langle A(\phi)\rangle=\langle \phi(x_1)\phi(x_2)\cdots\rangle$. Computing this average directly is hard because of the $Z_h^{-1}$ factor.
\subsubsection{Method of replicas}
In order to circumvent this problem, it is customary to utilize the method of replicas \cite{cardy_1996}. 
The idea is to consider $n$ copies of the theory, whose partition function is simply $Z_h^n$. 
From the $n\rightarrow 0$ limit we obtain the averaged free energy $\bar{\log Z}=\lim_{n\to 0}n^{-1}(\bar{Z^n}-1)$. 

Here we will present an alternative way to implement the replica method, more suitable to compute correlation functions \cite{Castellani_2005}.
For this let us insert $1=Z_h^{n-1}/Z_h^{n-1}$ inside the $h$-integral in \eqref{formula:1copy}. This gives
\be
\label{formula:1-repl}
\overline{\langle A(\phi)   \rangle }= \int \mathcal{D}h \, P(h) \ \frac{\int \mathcal{D} \vec \phi \ A(\phi_1)  \ e^{-\sum_{i=1}^n \mathcal  S[\phi_i,h]}}{Z_h^n} \ .
\ee
In the above  $\phi_i$ denotes $i$-th replica field, where $i=1,\dots , n$\,.
Notice that the r.h.s.~of this equation reproduces the l.h.s.~for any $n$, as long as it's a positive integer. If we can somehow analytically continue it to complex $n$, we may take the limit $n\to 0$, for which the denominator $Z_h^n\to 1$. Assuming that the limit exists and commutes with the integral over $h$, we  get the following:
\be
\label{form1-0}
\overline{\langle A(\phi)   \rangle }=\lim_{n\rightarrow 0}  \int \mathcal{D}h \, P(h)  \int \mathcal{D}\vec\phi \ A(\phi_1) \ e^{-\sum_{i=1}^n \mathcal S[\phi_i,h]}\ .
\ee

To make further progress, we assume that the magnetic field distribution is Gaussian, i.e. $P(h)\propto e^{-\frac 1{2 H} \int d^dx\,h(x)^2}$. Doing the path integral over $h$, we end up with a simple expression:
\be
\label{form1}
\overline{\langle A(\phi)   \rangle }=  \lim_{n\rightarrow 0}  \int \mathcal{D}\vec\phi \ A(\phi_1)  \ e^{- {\mathcal S}_n[\vec \phi]}\ ,
\ee
where the action is now free of disorder, however it contains a term which couples the $n$ replicas:
\be
\label{Sr}
\mathcal S_n[\vec \phi]= \int \dd^d x  \left[ \frac{1}{2}\sum_{i=1}^n(\partial \phi_i)^2 + \sum_{i=1}^nV(\phi_i) - \text{\small$\frac{1}{2}$}\kk\big(\sum_{i=1}^n  \phi_i\big)^2 \right] \ .
\ee
Equations \eqref{form1} and \eqref{Sr} define  averaged correlation functions.
In this formalism we can also access more observables, like random averages of products of correlation functions:
\begin{align}
\label{form2}
\overline{\la A(\phi)   \ra \la B(\phi)   \ra}= \lim\limits_{n\to 0}\int \mathcal{D}\vec\phi   \, A(\phi_1)  B(\phi_i)  \, e^{-{\mathcal S}_n[\vec \phi]}  \qquad (i\neq 1) \ .
\end{align}

To summarize, with the replica method, we obtained a prescription to compute observables in a disordered theory in terms of correlation functions of a quantum field theory of $n$ interacting fields. 
The price to pay is that we have to perform the subtle limit $n \rightarrow 0$.

\subsubsection{Cardy transformations}
The action \eqref{Sr} is invariant under the permutation of the $n$ replicas. However, in  \eqref{form1} and \eqref{form2}, the direction $\phi_1$ is treated differently from the directions $\phi_j$ for $j\neq 1$. 
This motivates the field redefinition introduced by Cardy \cite{CARDY1985123} (see also \cite{CARDY1985383}), that makes manifest only the permutation symmetry of  $n-1$ replicas and better captures the physics of the $n\to 0$ limit:
\be
\label{transf}
\varphi=\frac{1}{2}\big(\phi_1+\rho\big) 
\ , \qquad
\omega = \phi_1-\rho 
\ ,\qquad 
\chi_i =\phi_i - \rho\qquad(i=2,\ldots, n)\,, 
\ee
where we introduced the notation $\rho \equiv \frac{1}{n-1}\big(\phi_2+\cdots+\phi_n\big) $.
Note that only $n-2$ field $\chi_i$ are actually independent since $\sum_{i=2}^n \chi_i=0$.
When we use the transformation \eqref{transf} in the Lagrangian \eqref{Sr} we obtain a somewhat complicated result which can be schematically written as
\be\label{lagmain}
\mathcal S_n[\vec \phi]= \int d^d x  \Big[  \Lcal_{0} + \Lcal_{1}+  \Lcal_{2}  \Big]\ ,
\ee
where $ \Lcal_{0} $ is defined as 
\be
\label{Lrel}
\Lcal_{0} \equiv \partial \omega  \partial \varphi -  \frac{\kk}{2} \omega^2 +\omega V'(\varphi)+ 
\frac{1}{2} \sum _{i=2}^n \left[ (\partial \chi _i )^2 +\chi_i^2 V''(\varphi) \right] \ .
\ee
All the terms suppressed by some powers of $n$ are put in the Lagrangian $\Lcal_{2}$. All the remaining terms are assigned in the Lagrangian $\Lcal_{1}$. It can be checked that all these $\Lcal_{1}$ terms are less relevant than any term in the Lagrangian $ \Lcal_{0} $ (w.r.t.~dimension assignments which follow from $ \Lcal_{0} $'s kinetic term $\partial \omega  \partial \varphi -  \frac{\kk}{2} \omega^2$). In particular if one considers a potential $V$ such that  $\omega V'(\varphi), \chi_i^2 V''(\varphi)$ are marginal, then $\Lcal_{1}$ only contains irrelevant terms. 
For the sake of clarity we show here a few of the possible terms contained in $\Lcal_{1}$ and $\Lcal_{2}$:
\begin{align}
\Lcal_{1}& \supset	 V'''(\varphi) \times \Big\{ \sum_{i=2}^n \chi_i^3\,  ,  \ \sum_{i=2}^n \chi_i^2 \omega\, , \  \omega^3\,, \  \dots  \Big\} 
\label{LIrr}\ ,
\\
\Lcal_{2}& \supset n \left\{ (\partial \varphi )^2 \, , \  \kk \varphi \omega\, , \  V(\varphi) \, ,\  \dots \right\}
\label{Lsupp}
\ .
\end{align}
Following Cardy, in the present paper we will focus on the theory defined by $\Lcal_{0}$ and we will discard $\Lcal_{1}$ and $\Lcal_{2}$ terms. Naively this is legitimate, since they are either irrelevant or vanish as $n\to0$. One point of the companion paper \cite{paper2} will be to carefully analyze how the discarded terms affect the RG flow of the replicated theory, and to see if there is any subtlety.

\subsubsection{The emergence of Parisi-Sourlas supersymmetry}
In the following we want to show that, in the limit of zero replicas,  $\Lcal_{0}$ reduces to a Lagrangian with Parisi-Sourlas supersymmetry.

Taking the limit $n\rightarrow 0$ of the Lagrangian $\Lcal_{0}$ may not look straightforward since the dependence on $n$ appears through the fields  $\chi_i$, which are $n-2$ in number ($n-1$ fields subject to one constraint). 
However it is easy to show that in the limit of $n\rightarrow 0$ we can replace the $n-2$ fields $\chi_i$ with two anticommuting scalar fields $\psi$ and $\psib$:
\be
\label{fromchitopsi}
\frac{1}{2} \sum_{i=2}^n \chi_i [-\partial^2+V''(\varphi)] \chi_i \, \stackrel{n\to 0}{\longrightarrow} \, \psi [-\partial^2+V''(\varphi)] \bar{\psi} \ .
\ee
The proof of this statement consists in integrating out the  fields $\chi_i$ and  $\psi,\psib$ in their respective functional integrals and check that the result is the same. 
This manipulation is possible because the fields enter only quadratically in the Lagrangian.
The final result is that $\lim_{n\rightarrow 0}  \Lcal_{0}$ is equivalent to the theory defined by 
\be
\label{Lsusy}
\Lcal_{SUSY} = \partial^\m \omega  \partial_\m \varphi -  \frac{\kk}{2} \omega^2 +\omega V'(\varphi)+ 
  \partial^\m \psi\partial_\m \psib +\psi \psib  V''(\varphi)  \ .
\ee
This  is the celebrated Parisi-Sourlas Lagrangian, which surprisingly is invariant under super-Poincar\'e transformations (see section \ref{sec:susy}).
In order to make the supersymmetry manifest it is convenient to write the action in  superspace:
\be
\label{Ssusy}
\mathcal{S}_{SUSY}= 2\pi \int \dd^d x \dd \thetab \dd \theta \left[ -\frac{1}{2} \Phi \partial^a\partial_a \Phi + V(\Phi) \right] \, ,
\ee
where $\theta$ and $\thetab$ are scalar Grassmann coordinates. 
For later convenience we set $H=2$ and $\int  \dd \thetab \dd \theta \ \theta \thetab =1/(2\pi) $.\footnote{\label{note:norm}
If one wants to work with a general $H$, nice formulas will arise if one normalizes the Berezin integral as $\int  \dd \thetab \dd \theta \ \theta \thetab =H/(4\pi)$, and the superspace metric (see below) as  $x^2-\frac{4}{H}\theta \thetab$. The choice of $H$ can be thought of as a choice of units. Note that at the level of the SUSY Lagrangian \eqref{Lsusy} we can change $H$ by rescaling the fields, $\tilde\varphi = \a \, \varphi$, $\tilde \omega = \omega/\a$, $\tilde V(\tilde \phi)=\alpha^2 V(\phi)$, $\psi,\bar\psi=\text{inv}$ for some constant $\a$.
}
The super-Laplacian is defined in superspace as $ \partial^a\partial_a=\partial^2 +  2 \partial_{\thetab} \partial_{\theta}$.  
The superfield $\Phi(x,\theta,\thetab)$ is a function in superspace which can be expanded in components as follows:
\be
\label{def:Phi}
\Phi(x,\theta,\thetab)=\varphi(x) +\theta \psib(x)+ \thetab \psi(x) +\theta \thetab \omega(x) \ .
\ee
It is straightforward to check that by integrating out the variables $\theta,\thetab$ in  $\Scal_{SUSY}$, we recover the Lagrangian $\Lcal_{SUSY}$.

To summarize, we reviewed the logical path which brings one from the disordered theory \eqref{Sh} to the supersymmetric Parisi-Sourlas action \eqref{Ssusy}. This suggests that the IR fixed point of a random field model may be described by a supersymmetric theory. However, the path was long and involved some assumptions and approximations which are not obviously under control. These subtle issues are postponed to \cite{paper2} (where we will also review the vast literature and other approaches). In this paper we will study the SUSY theory \eqref{Ssusy} in its own right. 

\subsubsection{Relation between correlation functions}
Finally it is important to understand how the observables can be computed using the different actions.
Here, as an example, we focus on the two point function of $\phi$ of the random field theory \eqref{Sh}. Generalizations are straightforward.
Using Eqs. \eqref{form1} and \eqref{form2} we obtain two independent physical quantities:
\be
\label{2ptRF-0}
\overline{\la \phi(x_1)  \phi(x_2) \ra}=  \lim_{n\to0}\left\langle \phi_1(x_1)\phi_1(x_2)\right\rangle_n \ ,
\qquad 
\overline{\langle \phi(x_1) \rangle  \langle \phi(x_2) \rangle} = 
\lim_{n\to0}\left\langle \phi_1(x_1)\phi_i(x_2)\right\rangle _n\quad(i\ne 1)
\ .
\ee
where correlators in the r.h.s.~are the path integrals with the action $\Scal_n$, as in the r.h.s.~of \eqref{form1} and \eqref{form2}. We may simplify the second equation in \eqref{2ptRF-0} using the permutation symmetry of $\phi_{i\neq 1}$, by replacing $\phi_i\to \frac{1}{n-1}\sum_{i=2}^n \phi_i =\rho$. Further using the Cardy transformations \eqref{transf} to rewrite $\phi_1=\varphi+\frac{\omega}{2}$ and $\rho=\varphi-\frac{\omega}{2}$, we get
\be
\label{2ptRF-1}
\overline{\langle \phi(x_1) \rangle  \langle \phi(x_2) \rangle} = 
\left\langle(\varphi(x_1)+\tfrac 12\omega(x_1))(\varphi(x_2)-\tfrac 12\omega(x_2))\right\rangle 
\ ,
\ee
where the r.h.s.~can be now evaluated from the action $\mathcal{S}_{SUSY}$. The first equation in \eqref{2ptRF-0} can also be treated in a similar way. The resulting equations then involve the two point functions $\la \varphi\varphi \ra$, $\la \varphi\omega \ra$ and $\la \omega\omega \ra$. As we will explain in a moment, 
$\la \omega\omega \ra$ is zero if we assume supersymmetry, and then it can be dropped. Physically important quantities are the disconnected and connected correlation functions, which then acquire particularly simple expressions:
\be
\label{2ptRF}
\overline{\la \phi(x_1) \ra \la \phi(x_2) \ra}=  \left\langle \varphi(x_1)\varphi(x_2)\right\rangle \ ,
\qquad 
\overline{\langle \phi(x_1) \phi(x_2) \rangle -\langle \phi(x_1) \rangle  \langle \phi(x_2) \rangle} =  \left\la \varphi (x_1) \omega(x_2)  \right\ra
\ .
\ee

Now consider the two point function of the superfield $\Phi$ :
\be
\label{twoptsuperspace0}
\left\langle \Phi(x_1,\theta_1,\thetab_1) \Phi(x_2,\theta_2,\thetab_2))\right\rangle =F[(x_{1}-x_{2})^2-2(\theta_{1}-\theta_2)(\thetab_{1}-\thetab_2)]\,,
\ee
which by super-Poincar\'e invariance should be, as shown, some function of the superspace distance. This has several consequences. First
expanding in $\theta$'s we get 
\be
\label{twoptsuperspace00}
\left\langle \Phi(x_1,\theta_1,\thetab_1) \Phi(x_2,\theta_2,\thetab_2))\right\rangle =F[(x_1-x_2)^2]-2(\theta_{1}-\theta_2)(\thetab_{1}-\thetab_2) F' [(x_1-x_2)^2]\,.
\ee
Notice that the r.h.s. does not contain the quartic Grassmann term $\theta_1\thetab_1 \theta_2\thetab_2$. Via expansion \eqref{def:Phi}, the coefficient of that term would be a two point function $\langle \omega(x_1) \omega(x_2) \rangle$ which therefore vanishes by supersymmetry as claimed above.

To extract further consequences we set $x_2=\theta_2=\thetab_2=0$ and match with \eqref{def:Phi}; we get
\be
\label{twoptsuperspace}
\left\langle \Phi(x,\theta,\thetab)\Phi(0)\right\rangle= F(x^2-2\theta\thetab)= \left\langle \varphi(x)\varphi(0)\right\rangle +   \theta \thetab   \left\la \varphi (x) \omega(0)  \right\ra \, .
\ee
The two correlators in the r.h.s.~are precisely the two point functions of the random field theory \eqref{2ptRF}. Matching the coefficients of the $\theta$ expansion, we get
the following supersymmetric relation between them:
 \be
 \label{twoptsuperspace1}
  \left\la \varphi (x) \omega(0)  \right\ra= -\frac{1}{ r} \partial_r\left\langle \varphi(x)\varphi(0)\right\rangle\,,
 \ee
 where $r=\sqrt{x^\mu x_\mu}$.  
 Putting together \eqref{2ptRF} and \eqref{twoptsuperspace1} one gets a supersymmetric relation between connected and disconnected two point functions, which was checked by numerical studies in the RFIM in $d=5$ \cite{Picco3}. 
 We will come back to the two point functions of generic superfields in section \ref{Correlation_Functions_Embedding}.

\subsection{Parisi-Sourlas  supersymmetry and dimensional reduction}
\label{review:dim_red}
As we explained in the introduction, the Parisi-Sourlas supersymmetric theories \eqref{Ssusy} undergo dimensional reduction. 
We now review the original argument by Parisi and Sourlas \cite{Parisi:1979ka}.

The main claim is that the action \eqref{Ssusy} can be reduced to a $d-2$ dimensional action with no supersymmetry:
\be
\label{Sdm2}
\Scal_{red} =2\pi \int \dd^{d-2} \hat x \left[ -\frac{1}{2} \hat{\phi}  \, \partial^2 \, \hat \phi  + V(\hat \phi\,) \right] \, ,
\ee
where the potential $V$ is the same in the three formulations  \eqref{Sh}, \eqref{Ssusy} and \eqref{Sdm2}.
The reduction is in the sense that correlation functions of the  SUSY theory $\mathcal{S}_{SUSY}$ are equivalent to the ones of $\Scal_{red}$, when restricted to the submanifold defined by 
$ x^{d-1}=x^{d}=\theta= \thetab=0$. For example
\be\label{relcorr}
\langle \Phi(\hat x_1) \dots \Phi(\hat x_n) \rangle = \langle \hat \phi(\hat x_1)\dots \hat \phi(\hat x_n) \rangle\, ,
\ee
where $\hat x_i\in \mathbb{R}^{d-2}$. A similar relation should hold for correlation functions of composite operators built out of $\Phi$. 

Parisi-Sourlas proved the above correspondence perturbatively.  Their argument relied on this simple equality between integrals: 
\begin{align}
\label{RadialInt1}
\int  \dd^dx  \,  \dd \thetab \,  \dd \theta \, f(y^2) =  \int  \dd^{d-2} \, \hat x \,  f(\hat x^{2}) \, ,
\end{align}
where $y^2=x^2 -2 \theta \thetab$  is the norm (recall footnote \ref{note:norm})  of the superspace vector $y \in \mathbb{R}^{d|2}$ with $x\in \mathbb{R}^{d}$ (as we will review in the next section), while $\hat x^{2}$ is the norm of $\hat x \in \mathbb{R}^{d-2}$. For \eqref{RadialInt1} to hold, we assume $d\geq 2$ and that $f(r^2)$ decays faster than $r^{-(d-2)}$ at large $r$.\footnote{{To obtain \eqref{RadialInt1} one integrates out the Grassmann coordinates:
\be\int  \dd^dx  \,  \dd \thetab \,  \dd \theta \, f(y^2) = -\frac{1}{2\pi }\int d^dx \,  \frac{1}{r}\partial_{r}f(r^2) =-\frac{ \Omega_d}{2\pi}\int dr  \, r^{d-2} \, \partial_{r}f(r^2) = \Omega_{d-2} \int dr  \, r^{d-3}f(r^2)\,.  \ee Recall our non-standard normalization of the Berezin integral, footnote \ref{note:norm}. In the last steps we integrated by parts to simplify the radial integral, and reduced the angular part $\Omega_d=\frac{2\pi^{\frac{d}{2}}}{\Gamma(\frac{d}{2})}$ to $\Omega_{d-2}$. 
For $d=2$ the r.h.s. of \eqref{RadialInt1} simply becomes $f(0)$.}}
This argument can be extended to generic Feynman integrals thus obtaining a perturbative proof of dimensional reduction to all orders in perturbation theory. This is reviewed in appendix \ref{app:pertdimred}.

As is clear from \eqref{Ssusy} and \eqref{Sdm2}, the map of the two actions is quite simple. The functional forms of the two Lagrangians are the same, with the replacement $\Phi \to  {\hat \phi}$. This suggests a simple map between the operators of the two theories, for example
\be\label{simpmap}
\Phi^m \to {\hat \phi}^m\,.
\ee
As clear from \eqref{relcorr}, this map also dictates the dimensional reduction of correlation functions.

	\section{Non-perturbative Parisi-Sourlas superconformal symmetry}\label{sec:susy}
	In the previous section we reviewed how Parisi-Sourlas dimensional reduction arises. Usual arguments rely on the form of  the Lagrangian \eqref{Ssusy}. 
	For example the original proof of  \cite{Parisi:1979ka} was only perturbative. Later, some non-perturbative arguments, based on somewhat formal manipulations of the functional integral, appeared in \cite{CARDY1983470} and in \cite{KLEIN1983473} (the latter ones were put on a firmer mathematical ground in \cite{Klein:1984ff}).
As a downside they all rely on the form of the Lagrangian \eqref{Ssusy} and, strictly speaking, prove dimensional reductions only for $n$-point functions of the fundamental field $\phi$.
A different strategy to prove dimensional reduction was proposed by Zaboronsky \cite{Zaboronsky:1996qn} by means of  supersymmetric localization. This proof  clarifies which is the set of observables that undergoes dimensional reduction but still it relies on a Lagrangian formulation.

	In the following we pursue a different, axiomatic approach.
	This is at the foundation of the recent revival of the conformal bootstrap.
	In recent years a new way to compute CFT observables was discovered and applied to many important cases (for a review see \cite{Poland:2018epd}). The idea is that observables can be fixed by requiring that the theory satisfies very general axioms which arise from symmetries and the existence of an associative operator product expansion (OPE) in combination with extra assumptions like unitarity. This perspective has proven incredibly powerful to pinpoint CFTs and it suggests that this axiomatic point of view should be taken to define the CFTs  themselves. 
	The aim of this section is to show how a CFT with Parisi-Sourlas supersymmetry is defined in this  language. The consequences of dimensional reduction for CFTs will be addressed in section  \ref{sec:dimred}.\footnote{There is no relation between our work and the non-rigorous attempts to use the numerical conformal bootstrap for the problem of dimensional reduction by Hikami \cite{Hikami:2017sbg,Hikami:2018mrf}.}
	
	In the following subsection \ref{subsec:SymRep} we will focus on the symmetries and representations of Parisi-Sourlas theories.  In subsection \ref{subsec:SuperEmbeddingCorrelationFunctions} we will describe how to make the symmetries more manifest by means of an auxiliary embedding space. In this formalism it will be easy to describe correlation functions of the superconformal theory, as shown in subsection \ref{Correlation_Functions_Embedding}. The OPE and conformal blocks decomposition are finally discussed in subsection \ref{4pt_CBs}.
	\subsection{Symmetries and representations}
	\label{subsec:SymRep}
	The goal of this section is to describe the symmetries and associated representation of the IR fixed point of the Parisi-Sourlas theories. For pedagogical reasons, we first focus on  the symmetries of the theory along the RG flow. This allows us to carefully introduce the orthosymplectic group $OSp(d|2)$ and its representations which will play a crucial role in  the rest of the paper. In subsection \ref{subsec:symCFT} we proceed to analyze the Parisi-Sourlas superconformal fixed point, we illustrate the superconformal algebra and the representations under which the CFT operators transform.
	\subsubsection{The symmetries along the flow}
	\label{subsec:symQFT}
By construction, the action \eqref{Ssusy} enjoys super-Poincar\'e symmetry, namely it is invariant under supertranslations and superrotations. 
	Let us now forget about the action and describe the implications of the symmetries themselves.
	
	The supertranslations act as shifts in the superspace coordinates, both in the bosonic coordinates $x^\m$ and in the fermionic coordinates $\theta, \thetab$. It is convenient to introduce the superspace coordinate $y^a \equiv (x^\alpha,\theta,\thetab)$ where $a=1,\dots d, \theta, \thetab$ and $\alpha=1,\dots, d$ (unless explicitly said, we will consistently use lower case latin letters for superspace indices and greek letters for $\mathbb{R}^d$  indices). 
	Supertranslations, in superspace notation, take the form:
	\be
	y^a \to y'^a=y^a +c ^a \ ,
	\ee
	where $c^a$ are constants. Notice that  $c^\a$ are Grassmann-even (bosonic) while $c^\theta,c^\thetab$ Grassmann-odd (fermionic). The translation generators $P_a=(P_{\a},P_{\theta},P_{\thetab})$ act as derivatives $\partial_a\equiv \partial/\partial y^a$ in superspace. 
	$P_{\a}$ are bosonic generators, while $P_{\theta},P_{\thetab}$ are fermionic. They graded-commute as
	\be
	[P_a,P_b\}=0 \ .
	\ee
	As usual, the graded commutator $[X,Y\}$ is defined as the anticommutator $[X,Y\}=\{X,Y\}$ if both $X$ and $Y$ are Grassmann-odd and as the commutator $[X,Y\}=[X,Y]$ if one or both of them are Grassmann-even.

	The superrotations form  the orthosymplectic supergroup $OSp(d|2)$. This group will play a very important role in characterizing the little group of the superconformal field theory, thus defining the transformations of superprimary operators. For this reason in the following we carefully explain its action, algebra and representations.
	The orthosymplectic supergroup $OSp(d|2)$   is the group of super-linear transformations $y^a \to y'^a=R^a_{\ b} \, y^b$ that preserves the norm (recall that we set $H=2$)
	\be
	y^2\equiv y^a y_a =y^a g_{ab} y^b = x^2-2 \theta \bar \theta \,.
	\ee
	The orthosymplectic metric $g_{ab}$ is an invariant tensor of the group. It can be easily expressed in a block diagonal form, in terms of the $d$-dimensional flat metric and the symplectic metric:
	\be
	\label{metricOSp}
	g_{ab}\equiv 
	\left(
	\begin{array}{c c}
		I_{d} &0 \\
		0 &J_2 \\
	\end{array}
	\right) \ , \qquad 
	\ \ \ 
	I_{d}\equiv \mbox{diag}(\overbrace{1,1,\dots, 1}^{d}) \ , 
	\quad 
	J_2 \equiv \left(
	\begin{array}{c c}
		0 &-1 \\
		1 &0 \\
	\end{array}
	\right) \ .
	\ee
	
	The elements of the group $OSp(d|2)$ can be realized as $(d+2)\times (d+2)$ matrices $R^{a}{}_{b}$ which satisfy the following relation:
	\be
	R^{a}{}_{b} R^{c}{}_d\; g_{ac}=g_{bd} \ .
	\ee
	The matrices forming the supergroup can be written in block form as
	\be
	R^a{}_{b}= 
	\left(
	\begin{array}{c c}
		R^\alpha{}_\beta &R^\alpha{}_q\\
		R^p{}_{\beta} &R^p{}_q 
	\end{array}
	\right) \, ,
	\ee
	where we split indices $a=(\alpha,p)$, $b=(\beta,q)$ and $p,q\in {\theta,\thetab}$ run over the Grassmann directions.  
	The coefficients $R^\alpha{}_\beta$ and $R^p{}_q $ are even elements  of a Grassmann algebra, while  $R^\alpha{}_q$ and $R^p{}_{\beta} $ are odd elements of a Grassmann algebra.\footnote{
	We can consider a Grassmann algebra generated by an infinite set of anticommuting Grassmann variables $\theta_{1},\theta_{2},\ldots$
However this description is redundant. It suffices to consider 	$2d$ independent Grasmann variables to generate all the group elements. These are in one-to-one correspondence with the odd generators of the $OSp(d|2)$ algebra (see e.g. \cite{Hussin1994}). 
 }
	
	It is convenient to introduce the Lie algebra of $OSp(d|2)$, denoted as $\mathfrak{osp}(d|2)$ \cite{KAC19778, coulembier2012}. We name the generators of superrotations as $M^{a b}$, where $a,b=1,\dots,d,\theta,\thetab$. They are bosonic if both $a$ and $b$ have the same grading and fermionic otherwise. 
	$M^{a b}$ are graded antisymmetric, meaning that $M^{a b}=-(-1)^{[a][b]}M^{b a}$, where $[a]=0$ if for bosonic coordinates and  $[a]=1$ for fermionic ones (notice in particular that generators with two equal Grassmann indices $M^{\theta \theta}$ and $M^{\thetab\, \thetab}$ are non-zero, consistently with this equation). 
	They satisfy the following commutation relations:
	\be
	\label{algebraOSp(d|2)}
	[M^{a b},M^{c d}\}
	=g^{c b} M^{a d}-(-1)^{[a][b]}g^{c a} M^{b d}-(-1)^{[c][d]}g^{d b} M^{a c}+(-1)^{[a][b]+[c][d]}g^{d a} M^{b c} \,.
	\ee
	So far we encountered the generators $P_a$ which are in a vector representation of $OSp(d|2)$ and $M^{a b}$ which are in the graded antisymmetric. In the following it will be crucial to have under control more generic finite dimensional representations of the superrotations. To this end we now classify tensor representations of $OSp(d|2)$. They are associated to graded Young tableaux which take the following form \cite{Dondi:1980qs}:
	\be
	\label{OSpYT}
	{ \scriptsize
		\begin{ytableau}
			\, _{\cdots} &\, _{\cdots}&\, _{\cdots}&\, _{\cdots}&\, _{\cdots}&\, _{\cdots}&\\
			\resizebox{!}{0.3 cm}{\bf \vdots}&\resizebox{!}{0.3 cm}{\bf \vdots}& \resizebox{!}{0.3 cm}{\bf \vdots}& \resizebox{!}{0.3 cm}{\bf \vdots}& \raisebox{0.1 cm}{\resizebox{!}{0.12 cm}{\rotatebox[origin=c]{45}{\dots}}} \\
			\, _{\cdots} &\, _{\cdots}& \\
			\\
			\resizebox{!}{0.3 cm}{\bf \vdots} \\
			\\
		\end{ytableau}
	}\
	\raisebox{-0.35 cm}{
		$\Bigg\} [\frac{d}{2}]$
	} \ ,
	\ee
	where the $i$-th row has length $l_{i}$ and $l_{i}\geq l_{i+1}$. The first $ [\frac{d}{2}]$ rows can be arbitrarily long and they are roughly speaking associated to the $SO(d)$ part of the group. Similarly the first column may contain an arbitrary number of boxes, and it is related to $Sp(2)$.

	The graded Young tableaux is a straightforward generalization of the usual Young tableaux, with the following rules:
	\begin{itemize}
		\item Indices along the rows are graded symmetrized,
		\item Indices along the columns are graded antisymmetrized,
		\item All the super-traces are removed.
	\end{itemize}
		The idea can be illustrated with the following examples of tensors with two indices respectively belonging to the graded symmetric and the graded antisymmetric representations:
	\begin{align}
	{ \scriptsize
		\begin{ytableau}
		a & b \\
		\end{ytableau}
	} 
	\qquad 
	&t^{ab}=\frac{1}{2}\big(t^{ab}+(-1)^{[a][b]}t^{ba} \big) - {t^c}_ c \frac{g^{ab}}{d-2} \ ,
	\label{gradedsym}
	\\
		{ \scriptsize
		\begin{ytableau}
		a \\ 
		b \\ 
		\end{ytableau}
	} 
	\label{gradedantisym}
	\qquad  
	&t^{ab}=\frac{1}{2}\big(t^{ab}-(-1)^{[a][b]}t^{ba}\big)\,.
	\end{align}
	As already mentioned, the generators $M^{a b}$ belong to the second type \eqref{gradedantisym}.
	The action of the generators of superrotations on a tensor with $\ell$ indices takes the following form:
	\be
	\label{tensor_rot}
	[M^{a b},t^{a_1 \dots a_{\ell}} \}=\sum_{k=1}^{\ell} \s^{(k)}_{\{a_i\}} [\Sigma^{a b}]^{a_k}_{\phantom{a_k} c} \  t^{c \, a_1 \dots a_{k-1}  a_{k+1} \dots a_{\ell}} \ ,
	\qquad 
	[\Sigma^{a b}]^{d}_{\phantom{d} c}\equiv g^{d b} \delta^{a}_c -(-1)^{[a][b]} g^{d a} \delta_{c}^b \ ,
	\ee
	where $\s^{(k)}_{\{a_i\}}$ is the sign acquired by the tensor $t^{a_1 \dots a_{\ell}}$ after commuting $a_k$ to the left of $a_1$, namely $ t^{a_1 \dots a_{\ell}}=\s^{(k)}_{\{a_i\}} t^{a_k a_1 \dots a_{k-1}  a_{k+1} \dots a_{\ell}}$.
	For example $M^{a b}$ rotates the index of $P^c$ giving the following commutation relation:
	\be
	\label{MPcommutator}
	[M^{a b},P^{c}\} =g^{c b} P^{a}-(-1)^{[a][b]} g^{c a} P^{b} \ .
	\ee
	In the following it will be often convenient to use an index-free formalism. This can be achieved by contracting tensors with  polarization vectors of the form
	\be
	w^{a}\equiv (z^{\a},\zeta,\bar \zeta) \ ,
	\ee
	where $z^\a$ is a Grassmann even $d$-dimensional vector while $\zeta,\bar \zeta$ are Grassmann odd variables.
	For example, a graded symmetric tensor of spin $\ell$ is encoded by the following index-free notation:
	\be
	\label{tensortopoly}
	t(w) \equiv w_{a_\ell} \cdots w_{a_1} t^{a_1 \dots a_\ell}.
	\ee
	This is consistent since the polarization vectors graded-commute between themselves. A further simplification is to encode the super-tracelessness of the tensor by using polarization vectors which are null: $0=w^{a} w_{a}=z^2-2\zeta \bar \zeta$. The original tensor components can be extracted from $t(w)$ by taking derivatives:
	\be
	\label{polytotensor}
	t^{a_1 \dots a_\ell} = \frac{1}{\ell! (\frac{d}{2}-2)_\ell} D_w^{a_1} \cdots D_w^{a_\ell} t(w)\,,
	\ee
	where $D^a_w$ is an appropriate differential operator:
	\be
	\label{Tod}
	D_w^a=\big(\frac{d}{2}-2+w \cdot \partial_w\big) \partial_w^a - \frac{1}{2} w^a \partial^2_w \ .
	\ee
	Here the dot-product and the square are defined to build scalars of $OSp(d|2)$, e.g. $w_1 \cdot w_2 \equiv w_1^a \, w_{2 \, a}$ for some vectors $w_1, w_2$.  Of course to take derivatives we have to extend $t(w)$ away from the null cone $w^{a} w_{a}=0$, but the point of this construction is that the result does not depend on how the extension is taken and gives back the original graded symmetric, super-traceless tensor.
	This technology matches the one used in \cite{Costa:2011mg} for  $SO(d)$ tensors, besides that \eqref{polytotensor} and \eqref{Tod} depend on a shifted value of the dimension $d \to d-2$. 
	More general representations are handled by generalizing the strategy of \cite{Costa:2014rya, Costa:2016hju, Lauria:2018klo}: we contract the mixed symmetric tensors with a sequence of polarization vectors, where the indices of the $i$-th graded symmetric row of the Young tableau are contracted with polarization vectors $w_i$. We further ask for $w_i \cdot w_j =0$ to encode tracelessness. To undo the contractions and recover the original components, we apply derivatives $\partial^a_{w_i}$ and contract the resulting indices with opportune projectors. Interestingly the projectors for $OSp(d|2)$ take the same form as the ones of $SO(d)$ \cite{Costa:2016hju} after the shift $d \to d-2$ (one example of this is visible in \eqref{gradedsym}).

	Finally, let us mention that  the finite dimensional representations of $OSp(d|2)$ can be decomposed in terms of representations of $O(d)\times Sp(2)$. 
	The standard technique is explained in \cite{King:1975vf, Farmer:1983un}.

	\subsubsection{The symmetries of the fixed point}
	\label{subsec:symCFT}
	In this section we focus on the symmetries of fixed point of the RG flow of the action \eqref{Ssusy}. This is, by definition, invariant under superdilations. We will also assume that, along with superdilations, it is invariant under special superconformal transformations. Superconformal transformations are defined by analogy with conformal transformations as being locally given as a composition of a superrotation and a superdilation. The assumption of superconformal invariance can be justified by the usual arguments involving a traceless stress tensor, as for the usual scale vs conformal invariance story, and with usual caveats. We will see this later in this section. Superconformal invariance of free massless theory (i.e.~$V=0$ in \eqref{Ssusy}) can be checked explicitly \cite{KUPIAINEN1983380}. 
	
The generators of the transformations form the Parisi-Sourlas superconformal algebra, which is denoted by $\mathfrak{osp}(d+1,1|2)$. In this section we show some features of this algebra and its representations.\footnote{There is little literature about representations of this superconformal algebra,
	which violates spin statistics and thus cannot occur in unitary theories of interest to high energy physicists. The only relevant paper known to us is \cite{KUPIAINEN1983380}, which however treats explicitly only scalar superprimaries, and does not relate superconformal invariance to the super-stress tensor.} 	
	
	The algebra $\mathfrak{osp}(d+1,1|2)$ consists of the generators of superrotations $M^{ab}$, supertranslations $P^a$, superdilations $D$ and special superconformal transformations $K^a$, which correspond to the following superspace vector fields:
	\be
	\label{SCT_generators}
	P^a= \partial^a \, ,
	\qquad
	M^{ab}=y^a \partial^b - (-1)^{[a][b]} y^b \partial^a \, ,
	\qquad
	D=- y^a \partial_a \, ,
	\qquad
	K^a=2 y^a y^b \partial_b- y^b y_b \partial^a \, .
	\ee
	Their graded algebra takes the form:
	\begin{align}
	\label{susyalgebra}
	\begin{split}
	[D,P^{a}\} &=P^a\ ,
	\\  [D,K^{a}\} &=-K^a \ , \\
	[K^{a},P^{b}\}
	&=2 g^{ba} D - 2 M^{a b} \ ,
	\end{split}
	\end{align}
	where 	$g$ is the orthosymplectic metric \eqref{metricOSp}.  Here we omitted the commutator of superrotation generators which already appeared in \eqref{algebraOSp(d|2)} and  \eqref{MPcommutator} (the commutator of $M^{ab}$ and $K^a$ takes the same form as \eqref{MPcommutator}, since $K^a$ is also an $OSp(d|2)$ vector).
	All the other commutators vanish.
	
	The graded-commutators \eqref{susyalgebra} describe a ladder algebra with raising and lowering operators $P^a$ and $K^a$. 
	Superfields transform in the representation of the little group $OSp(d|2)\times U(1)$ generated by the superrotations $M^{ab}$ and superdilations $D$. A superprimary operator diagonalizes the superdilation operator and is annihilated by $K^a$:
	\be\label{KDaction}
	[K^a,\Ocal(0)\}=0 \ , \qquad [D,\Ocal(0)\}=\D \Ocal(x) \ .
	\ee
Since the condition for superprimaries includes the operator to be annihilated by the two Grassmann charges $K^{\theta},K^{\thetab}$, it is more restrictive that the condition for $SO(d+1,1)$ primaries which are only annihilated by $K^\alpha$. 
	Superrotations act on the operator $\Ocal(0)$ by rotating its  $OSp(d|2)$ indices according to \eqref{tensor_rot}. Finally $P^a$ creates superdescendants.
		
	So far we explained how to organize the operators of the theory in terms of superfields which transform naturally under the little group. Since every superconformal theory is also conformal, it is interesting to know how the spectrum splits in conformal multiplets of $SO(d+1,1)$.
		Let us show how this works for an operator $\Ocal$ with super-dimension $\D$ and transforming in a graded-commuting representation of spin $\ell$ of $OSp(d|2)$. First we expand the superfield in $\theta, \bar \theta$:
	\be\label{Oexpanded}
	\underbrace{\mathcal{O}^{a_1 \dots a_\ell }(x,\theta,\thetab)}_{\D}=
	\underbrace{\mathcal{O}^{a_1 \dots a_\ell }_{0}(x)}_{\D}+
	\theta \underbrace{ \mathcal{O}^{a_1 \dots a_\ell }_{\theta} (x)}_{\D+1}+
	\bar{\theta} \underbrace{ \mathcal{O}^{a_1 \dots a_\ell }_{\bar{\theta}}(x)}_{\D+1}+
	\theta\bar\theta \underbrace{ \mathcal{O}^{a_1 \dots a_\ell }_{\theta\bar\theta}(x)}_{\D+2}\,.
	\ee
	The four coefficients of the $\theta, \bar \theta$ polynomial, which we dubbed $\mathcal{O}_{0}, \mathcal{O}_{\theta}, \mathcal{O}_{\thetab}, \mathcal{O}_{\theta \thetab}$, have the same spin under $OSp(d|2)$, but they have shifted conformal dimensions, according to the mnemonic dimension rule that $[\theta]=[\bar \theta]=-1$.	The indices $a_i$ are naturally organized in finite irreducible representations of $OSp(d|2)$. We can further decide to exhibit the rotation symmetry of the CFT$_d$ by reorganizing the $OSp(d|2)$ representations in terms of the $SO(d)$ ones. E.g.~the zero component of a vector $\Ocal^a$ can be written as a collection of one $SO(d)$ vector (boson) and two scalars (fermions) as follows $\Ocal_0^a=(\Ocal_0^\a,\Ocal_0^{\theta},\Ocal_0^{\bar \theta})$. A similar decomposition holds for $\Ocal^a_\theta$, $\Ocal^a_{\bar \theta}$ except for the opposite statistics of vector and scalar components. 
Each of the four constituents $\mathcal{O}_I^{a_1 \dots a_\ell }$ ($I=0,\theta,\bar\theta, \theta\bar\theta$) of a general graded-symmetric $OSp(d|2)$ spin $\ell\ge 2$ tensor superprimary multiplet would thus give rise to four conformal primaries $\Ocal_I^{\m_1 \dots \m_\ell}$, $\Ocal^{\m_1  \dots \m_{\ell-1}\theta}_I$, $\Ocal^{\m_1  \dots \m_{\ell-1}\bar\theta}_I$ and $\Ocal^{\m_1  \dots \m_{\ell-2}\theta \bar\theta}_I$ of decreasing $SO(d)$ spins $\ell$, $\ell-1$, $\ell-1$ and $\ell-2$ and of accordingly alternating statistics.\footnote{Since we are assuming graded symmetry, if we set two indices to the same Grassmann value we obtain zero, e.g. $\Ocal^{\m_1  \dots \m_{\ell-2}\theta \theta}_I=0$.} 
	
	Let us now discuss some features of short superconformal multiplets.
	To begin, we shall see how shortening conditions are used to fix the scaling dimension of the supermultiplet.
	As a first example we consider the free scalar superprimary $|\Ocal\rangle\equiv \Ocal(y=0) |0\rangle$, defined by $P^a P_a |\Ocal\rangle = 0$. 
	By asking that its norm vanishes and using the commutation relations \eqref{susyalgebra}, we discover that the dimension of $\Ocal$ is fixed as $d/2-2$.
	We can similarly consider a conserved superprimary operator $\Ocal^{a_1\dots a_\ell}$, in the graded symmetric representation of $OSp(d|2)$. We require that the norm of $ P^a |\Ocal_a^{\phantom{a}a_2\dots a_\ell}\rangle$ vanishes. This then fixes its superconformal dimensions to $\D=\ell+d-4$. As an example, when $\ell=1$, we have 
	\be
	0=\langle \Ocal^b| K_b P^a |\Ocal_a\rangle=\langle \Ocal^b| [K_b,P^a\} |\Ocal_a\rangle=2(\D-d+3)\langle \Ocal^a|\Ocal_a\rangle \, .
	\ee
Here we defined a conjugate state by $| \Ocal\rangle ^\dag \equiv\langle\Ocal|\equiv \lim_{y\to 0} |y|^{- 2\D}  \langle 0| \Ocal (y^a/y^2)$ (when an operator has indices they have to be contracted with reflection matrices $g^{ba}-2y^ay^b/y^2$). Since the conjugation is defined by superspace inversion, we can apply the usual logic and obtain that $P_a^\dag=K_a$ ---indeed the generator of special superconfrormal transformations is obtained from the successive application of inversion-translation-inversion in superspace.
	
	We continue our discussion by showing how short CFT$_d$ multiplets are embedded into the short superconformal multiplets of the Parisi-Sourlas CFT.
	Let us exemplify how this works for a spin-two superprimary $\Tcal^{a b}$ (generalizations to other short multiplets are straightforward), which satisfies the conservation equation:
	\be\label{supcons}
	\partial^a \mathcal{T}_{a b}=0\,.
	\ee
	We recall that  graded symmetry and the super-tracelessness respectively imply  $\Tcal_{ab}=(-1)^{[a][b]}\Tcal_{ba}$  and $g^{ab}\Tcal_{ab}=0$. Furthermore, as we showed above, conservation \eqref{supcons} fixes the dimension of $\Tcal^{a b}$ as $d-2$.
	We refer to $\Tcal^{a b}$ as the \emph{super-stress tensor} of the Parisi-Sourlas CFT, associated with the  supertranslation symmetry. Of course since every Parisi-Sourlas CFT is also a usual CFT$_{d}$, as such it should contain the usual stress tensor $T^{\a \b}$ primary, of dimension $d$. It would be natural to guess that $T^{\a \b}$ is contained in the $\Tcal^{a b}$ supermultiplet, and indeed that's what happens.
		Expanding $\mathcal{T}_{ab }(x,\theta)$ as in \eqref{Oexpanded} we get
	\be
	\label{Texpanded}
	\underbrace{\mathcal{T}^{a b}(x,\theta,\thetab)}_{d-2}=
	\underbrace{\mathcal{T}^{a b }_{0}(x)}_{d-2}+
	\theta \underbrace{ \mathcal{T}^{a b}_{\theta} (x)}_{d-1}+
	\bar{\theta} \underbrace{ \mathcal{T}^{a b}_{\bar{\theta}}(x)}_{d-1}+
	\theta\bar\theta \underbrace{ \mathcal{T}^{a b}_{\theta\bar\theta}(x)}_{d}\,.
	\ee
	By dimensionality one can guess that the usual $d-$dimensional stress tensor $T^{\a\b}=\mathcal{T}^{\a \b}_{\theta\bar\theta}$, the spin-two part of the highest-dimension constituent $\mathcal{T}^{a b}_{\theta\bar\theta}$. As a check let us verify
that equation \eqref{supcons} implies the conservation of $\mathcal{T}_{\theta \thetab }^{\a \b}$.
	In fact by plugging \eqref{Texpanded} in \eqref{supcons} and taking the $\theta\bar{\theta}$ coefficient, one gets precisely
	\be
	\partial_\a \mathcal{T}_{\theta \thetab }^{\a \b}(x)=0\,,
	\ee
	where $\a,\b$ run from $1$ to $d$\,.\footnote{Grassmann derivatives like $\partial_\theta \mathcal{T}^{\theta \b}$ cannot contribute to this equation because they can never produce a $\theta \bar\theta$ term.} 
	Since $ \mathcal{T}_{\theta \thetab }^{\a\b}$ is a conserved spin-two tensor of $SO(d)$, with conformal dimension $d$, we conclude that it should correspond to the $d$-dimensional stress tensor.
	It is instructive to explicitly check this statement in free theory as we show in appendix \ref{Free_ex}.
	
The super stress tensor is a very important operator since many properties of a CFT descend from its existence.
We shall define locality of the Parisi-Sourlas theory by the following equation:
	\be\label{Tlocal}
	\int_{V } \dd^d x  \dd \theta\dd \thetab \, \partial^a \mathcal{T}_{ab}(y)\Ocal(0)=\partial_b \Ocal(0)\, ,
	\ee
	where the l.h.s.~is an integral over a region  $V \subset \mathbb{R}^{d|2}$ that contains the point $y=0$ and $\Ocal$ is local operator which, for simplicity, we consider to be a scalar.\footnote{
The superspace topology is defined by its reduced bosonic topology. The mnemonic rule being that Grassmann variables are much smaller than the bosonic ones, since they square to zero. E.g. the interior of a superball is identified by the interior of the ball obtained by setting its Grassmann coordinates to zero.
See  \cite{Witten:2012bg} for a review on topology and integration over supermanifolds.   
	} 
	By using Stokes' theorem (see e.g. \cite{Witten:2012bg}) the integral in  \eqref{Tlocal} can be written as a surface integral on the boundary  $\partial V$  of the region $V$.
	Because of the conservation of the stress tensor, the  l.h.s.~of \eqref{Tlocal} defines a topological surface operator, which does not depend on the shape of the surface $\partial V $ (as long as other operator insertions are not crossed).
	We can e.g.~consider $V \in \mathbb{R}^{d|2}$ to be a superball of radius $\varrho$, its boundary a supersphere.  Since $\varrho$ can be taken arbitrarily small, we expect the r.h.s.~to be written as a local operator inserted at the origin. Moreover, since the super-stress tensor generates supertranslations, the r.h.s.~is fixed in terms of a superspace derivative acting on $\Ocal$. Eq.~\eqref{Tlocal} also fixes the normalization of the super-stress tensor.

	Finally let us come back to the assumption of superconformal invariance.
	In usual CFTs, conformal invariance descends from the condition of tracelessness of the stress tensor. Dilation invariance of the fixed point imposes that the stress tensor is either traceless or it satisfies $T^\m_{\phantom{\m}\m}=\partial^\m V_\m$ for some vector field $V_\m$ called ``virial current''. Conformal invariance follows from scale invariance if the virial current is absent (or it is a descendant of some spin-two operator $W_{\m \rho}$, namely $V_\m= \partial^\rho W_{\m \rho}$) \cite{Polchinski:1987dy}. In practice we are trading the assumption of conformal invariance with the assumption of the absence of a virial current. The latter is clearly more satisfying since we do not expect to have, in a generic interacting CFT, a non-conserved vector operator with dimension $d-1$. This expectation should hold in unitary and non-unitary theories alike. 
	
	The same kind of logic can be adopted in our supersymmetric setup. Indeed the absence of a super-virial current $\mathcal{V}^a$---non-conserved and with superdimension $(d-3)$---and the presence of superdilations are enough to impose that the super-stress tensor is traceless and therefore to establish superconformal symmetry.\footnote{Tracelessness of the super-stress tensor is equivalent to the invariance under local rescaling of  the superspace metric (near flat space). The latter is sufficient to reconstruct all the superconformal Killing vectors  including the ones associated to special superconformal transformations. As in usual CFTs, one can then further define additional topological surface operators integrating $\mathcal{T}_{ab}(y)$ of \eqref{Tlocal} against the superconformal Killing vectors $\e^b(y)$, thus recovering all generators of \eqref{SCT_generators}.
}

	\subsection{Super-embedding space and correlation functions}
	\label{subsec:SuperEmbeddingCorrelationFunctions}
	In this section we analyze the correlation functions of the Parisi-Sourlas CFT.
	We first introduce the super-embedding space formalism, which generalizes the usual embedding space formalisms of \cite{Costa:2011mg}. This simplifies the action of conformal transformations and trivializes the problem of finding conformal invariant combinations to build correlation functions.
	We then apply this technique to correlation functions and give a few examples of how they are constrained by supersymmetry. Finally we focus on the OPE and on how this can be used to decompose four point functions in superconformal blocks.
	\subsubsection{Superembedding space}
		The Parisi-Sourlas CFT is invariant under the orthosymplectic group $OSp(d+1,1|2)$, which does not act linearly in superspace $\mathbb{R}^{d|2}$. The idea of the super-embedding formalism is to uplift the theory to the space $\mathbb{R}^{1,d+1|2}$, where the group acts linearly. 	It is then possible to go back to the usual superspace $\mathbb{R}^{d|2}$ by using a simple projection.

	Given a point $P^M=(P^0,P^{\m},P^{d+1},\theta,\bar \theta)$ in the super-embedding space $\mathbb{R}^{1,d+1|2}$, its norm is defined as 
	\begin{align}
	P^2\equiv P^M P_M \equiv P_M \bar g^{MN} P_N =-(P^0)^2+(P^\m)^2+(P^{d+1})^2+2 \, \thetab  \theta   \ ,
	\end{align}
	where the metric $\bar g$ is a block diagonal matrix build in terms of two pieces: the flat metric $I_{1,d+1}$ of $\mathbb{R}^{1,d+1}$ and the symplectic metric $J_2$ of $Sp(2)$:
	\be
	\label{s_embedding_metric}
	\bar g_{AB}= 
	\left(
	\begin{array}{c c}
		I_{1,d+1} &0 \\
		0 &J_2 \\
	\end{array}
	\right) \, ,
	\qquad \mbox{where}
	\qquad 
	I_{1,d+1}=\mbox{diag}(\underbrace{-1,1,\dots, 1}_{d+2}) \ ,
	\quad 
	J_2=\left(
	\begin{array}{c c}
		0 &-1 \\
		1 &0 \\
	\end{array}
	\right) \, .
	\ee
	In order to get rid of the two extra dimensions we consider only the points $P^M$ which belong to the projective null cone $P ^2=0$ with $P \sim \l P$.
	To get back a result parametrized by the coordinates $(x,\theta,\bar\theta)$ in physical space, we will restrict each point $P^A$ to lie in the super-Poincar\'e (sP) section:
	\be
	P^A \big |_{\textrm{sP}} =\left(\frac{1+x^2+2 \thetab  \theta}{2},x^{\m},\frac{1-x^2- 2 \thetab  \theta}{2},\theta, \thetab \right) \, .
	\ee
	Scalar primary superfields extended to the projective null cone satisfy the following homogeneity condition:
	\be
	\label{scalingO}
	\Ocal(\l P)=\l^{- \D}\Ocal(P) \, ,
	\ee
	for any $\l \in \mathbb{R}$.
	We can further uplift operators that transform as $OSp(d|2)$ tensors, by imposing \eqref{scalingO} and by allowing their indices to take values $A=0,\dots,d+1,\theta,\thetab$.
To get a description which is not redundant, the operators are required to satisfy the transversality condition $P_{A_1} \Ocal^{A_1 A_2 \dots A_\ell}(P)=0$. We also have a condition that ``pure gauge'' tensors proportional to $P_{A_i}$ project to zero. To deal with these constraints, 
	it is convenient to contract the tensor indices with super-embedding space polarization vectors. For example for a spin $\ell$ graded symmetric superfield we define
	\be
	\Ocal(P,Z)\equiv \Ocal^{A_1 \dots A_\ell}(P) \ Z_{A_\ell} \cdots Z_{A_1} \, ,
	\ee
	where  polarization vectors are chosen as $Z^M=(Z^0,z^\m,Z^{d+1},\z,\bar \z)$ and $\z,\bar \z$ are Grassmann odd coordinates while $Z^0,z^\m,Z^{d+1}$ are Grassmann even and $z^\m$ is a $d-$dimensional vector. The transversality condition and the super-tracelessness conditions can be encoded by choosing polarization vectors that respectively satisfy $P \cdot Z = 0 $ and  $Z^2=0 $.
	When we restrict coordinates to the super-Poincar\'e section, the polarization vectors are projected to vectors tangent to this section, as follows:
	\be
	\label{ZtosP}
	Z^M  \big |_{\textrm{sP}} =\left((x\cdot z+ \z \bar \theta + \bar \z \theta) , z^\m, - (x\cdot z+ \z \bar \theta + \bar \z \theta), \z, \bar \z \right) \, .
	\ee
	Notice that the condition $Z^2=0$ is projected to $Z^2  \big |_{\textrm{sP}}= z^2 -  2 \z  \bar \z=0$ in physical space, which indeed encodes the super-tracelessness condition of $OSp(d|2)$ tensors as explained below \eqref{tensortopoly}. One can also consider more generic representations for the superfields $\Ocal$, analogously to what was done in \cite{Costa:2014rya} for the usual conformal group. Tensors of $OSp(d|2)$ can be associated to Young tableaux of the form \eqref{OSpYT}.
	One can contract the indices of each row with a polarization vector $Z^{(i)}$ such that $P \cdot Z^{(i)} = 0 $. It is possible to choose $Z^{(i)} \cdot Z^{(j)} = 0 $. Finally one can project them into the super-Poincar\'e section in terms of physical space polarization vectors $z^{(i)},\z^{(i)}, \bar {\z}{}^{(i)}$ by a straightforward generalization of \eqref{ZtosP}.
	
	In super-embedding space, the generators  of infinitesimal orthosymplectic transformations $L_{AB}$ take a particularly simple form:
	\be
	\label{def:LAB}
	L_{AB} \equiv P_{A} \partial_{P_{B}} -(-1)^{[A][B]} P_{B} \partial_{P_{A}} \ .
	\ee
	$L_{AB}$ packages together supertranslation, superrotations, superdilations and special superconformal  transformations, while making the $OSp(d+1,1|2)$ action linear, and the invariance manifest.
	Indeed it is easy to see that $L_{AB}$ satisfies the $\mathfrak{osp}(d+1,1|2)$ algebra.
	In this notation the quadratic superconformal Casimir takes a very compact form:
	\be
	\label{quadCas}
	\frac{1}{2} \bar g^{AD}\bar g^{BC} L_{AB} L_{CD} \ .
	\ee
	The eigenvalue of the Casimir on any operator within a superconformal multiplet labelled by the super-dimension $\D$ and the $OSp(d|2)$ spin $\ell$ is 
	\be 
	\label{EigenvalueCasimir}
	c^{d|2}_{\D,\ell}= \D \big(\D-(d-2) \big)+ \ell \big(\ell+(d-2)-2\big) \ .
	\ee
	This can be obtained by a direct computation as we explain in section \ref{CB_red}. 
	Notice that this eigenvalue exactly equals the Casimir eigenvalue for a spin $\ell$, dimension $\Delta$ primary in a non-supersymmetric CFT living in $d-2$ dimensions. This property is indeed true also for generic representations.
	This fact will be crucial for the dimensional reduction to work.
	\subsubsection{Correlation functions in super-embedding space}
	\label{Correlation_Functions_Embedding}
	Using the super-embedding formalism it is easy to classify superconformal invariants which in turn can be used to write correlation functions. 
	In the following we show how to use this formalism in some simple examples. 
	It will be clear that all applications are straightforward generalizations of what happens in usual CFTs as detailed in \cite{Costa:2011mg}.
	
	Since $OSp(d+1,1|2)$ transformations act as linear transformations of the points $P^A$, conformal invariants are just scalar products $P_i \cdot P_j$. 
	For example a two point function of a scalar operator $\Ocal$ with superconformal dimension $\D$ can be written as
	\be
	\label{2ptgen}
	\la \Ocal(P_1)\Ocal(P_2) \ra =\frac{1}{(P_{12})^{\D}} \ ,
	\ee
	where the scaling is fixed by the requirement \eqref{scalingO} and we defined $P_{ij} \equiv - 2 P_i \cdot P_j$ where $-2$ is a convenient normalization.
This restricts to the super-Poincar\'e  section as
	\begin{align}
	-2 P_{i} \cdot P_{j} \big |_{\textrm{sP}} 
		&
	={y^2_{ij}} = (x_{i j})^2 -2 \theta_{ij} \thetab_{ij} \ ,
	\end{align}
	where $\theta_{ij}\equiv \theta_i - \theta_j$ , $\thetab_{ij}\equiv \thetab_i - \thetab_j$.
	Thus restricting \eqref{2ptgen} to the Poincar\'e section, we get the two point function in superspace: 
	By further expanding the two point function as a polynomial in the Grassmann coordinates we obtain
	\be
	\la \Ocal(P_1)\Ocal(P_2) \ra \big |_{\textrm{sP}}  =\frac{1}{(x_{12}-2 \theta_{12} \thetab_{12} )^{2\D}}\,.
	\ee
	This is similar to Eq.~\eqref{twoptsuperspace0} from the previous section for the fundamental superfield $\Phi$ of the Parisi-Sourlas Lagrangian. In particular we extract correlation functions between constituents of the $\theta$ expansion of $\Ocal$:
	\be
	\la \Ocal_0(x_1) \Ocal_0(x_2)\ra=\frac{1}{(x_{12})^{2\D}}\,,\qquad 
	\la \Ocal_0(x_1) \Ocal_{\theta\thetab}(x_2)\ra=  \frac{2 \D}{(x_{i j})^{2(\D+1)}} \ ,
	\ee
	in agreement with \eqref{twoptsuperspace1}, while the highest dimension constituent $\Ocal_{\theta\thetab}$ has zero two point function by the argument for the vanishing $\la \omega\omega\ra$ given there:\footnote{
		In fact this is a generic property of any $n$-point function: the highest component, proportional to $\theta_1\thetab_1\cdots \theta_n\thetab_n$, has to vanish. This is a simple consequence of supertranslation invariance, which can be used to translate any $\theta_i$ to zero.
	}
	\be
	\la \mathcal{O}_{\theta\bar\theta}(x_1) \mathcal{O}_{\theta\bar\theta}(x_2) \ra =0 \ .
	\ee
		It is important to stress that one should not consider $\omega$ or more generally any $\mathcal{O}_{\theta\bar\theta}$ as null states which can be modded out from the theory, in fact their correlation function with other operators may be non-zero, as is evidenced already by the non-zero two point function $\la \omega(x) \varphi(y) \ra\neq 0$.

	We can also consider two point functions of operators $\Ocal$ transforming in non-trivial representation of $OSp(d|2)$. For example if $\Ocal$ is a graded symmetric superprimary with super-dimension $\D$ and spin $\ell$, we have
	\be
	\la \Ocal(P_1,Z_1)\Ocal(P_2,Z_2)= \frac{(H_{12})^{\ell}}{(P_{12})^{\D}} \, ,
	\ee
	where the denominator is fixed by  \eqref{scalingO}  and $H_{12}$ is the unique conformal invariant which does not scale in $P_i$ and can be built out of null and transverse polarization vectors $Z_i$ (see \cite{Costa:2011mg}):
	\be
	H_{ij}\equiv \frac{(Z_i \cdot Z_j) (P_i \cdot P_j) - (Z_i \cdot P_j) (Z_i \cdot P_j) }{P_i \cdot P_j}\, .
	\ee
	More complicated representations can be taken into account generalizing the formalism of \cite{Costa:2014rya, Costa:2016hju, Lauria:2018klo}.
	
	Let us see how to apply this technology to three point functions. E.g.~take two scalar operators $\Ocal_1,\Ocal_2$ with dimensions $\D_1,\D_2$ and a spin $\ell$ operator $ \Ocal_3$ dimensions $\D_3$. Their three point function is written as
	\be\label{3ptfn}
	\la \Ocal_1(P_1) \Ocal_2(P_2) \Ocal_3(P_3,Z_3) \ra = \frac{\l_{123} \ V^\ell_{3,12} }{(P_{12})^{\frac{\D_1+\D_2-\D_3}{2}}(P_{23})^{\frac{\D_2+\D_3-\D_1}{2}}(P_{13})^{\frac{\D_3+\D_1-\D_2}{2}}} \ ,
	\ee
	where $\l_{123} $ is the associated OPE coefficient. The powers of $P_{ij}$ are fixed by the requirement  \eqref{scalingO}. The term $V_{3,12}$ does not scale in $P_i$ and it is the only conformal invariant which can be built out of the null and transverse polarization vector $Z_3$ \cite{Costa:2011mg}:
	\be
	V_{i,jk} \equiv \frac{(Z_i \cdot P_j) (P_i \cdot P_k) -(Z_i \cdot P_k) (P_i \cdot P_j)}{\sqrt{(P_{j} \cdot P_{k})(P_{j}\cdot P_{i})(P_{k}\cdot P_i)}} \ .
	\ee
	More complicated cases are again handled by following e.g. \cite{Lauria:2018klo}.

		Let us then consider a four point function of scalar superprimaries $\Ocal_{i}$ with dimensions $\D_i$ :
	\be
	\label{def:4pt}
	\la \Ocal_1(P_1) \Ocal_2(P_2) \Ocal_3(P_3) \Ocal_4(P_4) \ra =\Kcal(P_i) f(U,V) \, ,
	\qquad 
	\Kcal(P_i)
	\equiv
	\frac{ 
		\left(\frac{P_{24}}{P_{14}} \right)^{\frac{\D_1-\D_2}{2}}  \left(\frac{P_{14}}{P_{13}} \right)^{\frac{\D_3-\D_4}{2}}}{(P_{12})^{\frac{\D_1+\D_2}{2}}(P_{34})^{\frac{\D_3+\D_4}{2}}}
	\, ,
	\ee
	where $\Kcal(P_i)$ is a kinematic function that takes into account the correct scaling required by \eqref{scalingO}. The result is fixed up to a function $f$ of the two super-cross-ratios  which are scaleless in $P_i$ :
	\be
	\label{supercrossratios}
	U\equiv \frac{P_{12} \, P_{34} }{ P_{13} \, P_{24} }
	\ ,
	\qquad
	V \equiv \frac{P_{14} \, P_{23} }{ P_{13} \, P_{24} }
	\ .
	\ee
	Finally we can apply the same reasoning to a four point function of spinning operators $\mathcal{O}_i(P_i,Z_i)$. In super-embedding space, the conformal invariants are the same as the ones introduced by  \cite{Costa:2011mg}, namely one needs combinations $P_i \cdot P_j$ and $P_i \cdot Z_j$ which  scale opportunely in the $P_i,Z_i$ and which are transverse. 
	
	The reader used to the formalism introduced in \cite{Costa:2011mg}, may have recognized that the super-embedding formalism is a straightforward generalization. So we avoid further details. 
	\subsubsection{OPE and conformal blocks decomposition}
	\label{4pt_CBs}
	In this section we exemplify how the OPE works in a Parisi-Sourlas CFT. We then apply the OPE to a four point function to obtain its superconformal block decomposition.
	
	The OPE can be conveniently expressed in superspace. For simplicity we focus on the OPE between two scalar superfields $\Ocal_1,\Ocal_2$ with dimensions $\D_1,\D_2$:
	\be
	\label{super_OPE}
	\Ocal_1(y)  \Ocal_2(0) =  \sum_{\D,\ell}  \frac{\l_{12\Ocal}}{|y|^{\D_1+\D_2-\D+\ell}} \ [ y_{a_\ell} \cdots y_{a_1} \Ocal^{a_1 \dots a_\ell}(0) + \text{superdesc.} ]\ ,
	\ee
	where the sum runs over the possible graded symmetric tensor superprimaries $\Ocal$ with dimension $\D$ and $OSp(d|2)$-spin $\ell$. Indeed since $\Ocal_1,\Ocal_2$ are scalars, the indices of the exchanged operators can only be contracted with the superspace points $y^a$, which graded commute between themselves, so only graded symmetric primaries contribute.
	The `superdesc.' in \eqref{super_OPE} denotes the contribution of the superdescendants, i.e.~superspace derivatives ($\partial^a$ and higher order) of the superprimaries. As usual, these terms are completely fixed by superconformal invariance. Quantities not fixed by the symmetry are the dimensions of fields and the coefficients $\l_{12\Ocal}$, which already appeared in the three point function \eqref{3ptfn}.
	
	By taking the OPE of operators $12$ and $34$ inside the four point function \eqref{def:4pt}, we can express the function $f$ as a sum over superconformal blocks $g^{d|2}$ :
	\be
	f(U,V) = \sum_{\D,\ell} \l_{12\Ocal} \l_{\Ocal34} \ g^{d|2}_{\D,\ell}(U,V) \, ,
	\ee
	where, as in \eqref{super_OPE}, the sum runs over superprimaries $\Ocal$ labelled by  $\D,\ell$ and $ \l_{\Ocal34}$ are the OPE coefficients of the right channel $34$.
	Each conformal block resums the contribution of the  primary $\Ocal$ and all its descendants into a single function of the super-cross-ratios  \eqref{supercrossratios}.
	Even if the superblocks can be defined by applying the left and right OPEs to a two point function, there are more convenient techniques to compute them.
	One of the most efficient ones is the super-analogue of \cite{Dolan:2003hv}---to find them via the eigenfunctions of a second-order partial differential operator, the super-Casimir $C$:
	\be
	\label{casPDE}
	C \; G^{d|2}_{\D,\ell}(P_i)= c^{d|2}_{\D,\ell}\ G^{d|2}_{\D,\ell}(P_i) \, ,
	\qquad 
	\qquad 
	G^{d|2}_{\D,\ell}(P_i)\equiv\Kcal(P_i) g^{d|2}_{\D,\ell}(U,V) 
	\, .
	\ee
	The eigenfunctions $G$ are sometimes called conformal partial wave and are related to the conformal blocks $g$ by the kinematic factor \eqref{def:4pt}. The eigenvalue $c^{d|2}_{\D,\ell}$ of the Casimir equation is given by \eqref{EigenvalueCasimir} and the super-Casimir differential operator $C$ takes the form:
	\be
	\label{Casimir_Op}
	C\equiv \frac{1}{2} \bar g^{AD}\bar g^{BC} (L_1+L_2)_{AB}(L_1+L_2)_{CD} \ ,
	\ee
	where $L_i$ are the generators of rotations \eqref{def:LAB} acting on points $P_i$. 
	This discussion can be easily extended to four point functions of non-scalar operators by considering  generators of the rotations $L_i$ which also act on the polarization vectors of the external operators. 
	More generic exchanged operators can be taken into account by considering the opportune eigenvalues.
	In section \ref{CB_red} we will show how to compute the superconformal blocks in terms of conformal blocks of a non-supersymmetric CFT.
	
	\section{Dimensional reduction }\label{sec:dimred}
	In this section we want to show how dimensional reduction \cite{Parisi:1979ka, CARDY1983470, KLEIN1983473, Klein:1984ff, Zaboronsky:1996qn} works in the context of axiomatic CFTs.
	In particular we want to clarify how the Parisi-Sourlas CFT is related to a CFT that lives in $d-2$ dimensions. 
	
	In the next section we define what we mean by dimensional reduction.
	Then we explain that the locality of the Parisi-Sourlas CFT implies the locality of the theory in $d-2$.
	We further demonstrate how dimensional reduction works by showing that the superconformal blocks of the Parisi-Sourlas CFT match the conformal blocks of a $d-2$ dimensional CFT. Finally we describe how the equality of blocks implies a neat relation between usual conformal blocks in $d$ and $d-2$ dimensions.
	\subsection{Dimensional reduction and correlation functions}\label{dimredsec}
		The basic idea of dimensional reduction is that certain correlation functions of the Parisi-Sourlas CFT restricted to the subspace $\mathbb{R}^{d-2}\subset \mathbb{R}^{d}$ become correlation functions of a $d-2$ dimensional CFT without supersymmetry. 
	Of course any CFT restricted to a subspace defines a set of conformal correlators living in that subspace. This follows simply from the fact that $d$-dimensional conformal group restricts to the conformal group of the subspace. 
	What is unusual and non-trivial about the Parisi-Sourlas dimensional reduction is that, unlike a simple restriction, it produces a local theory. Before we explain this, let us recall some basic features of usual restrictions.

	A ``restriction'' is what happens to a CFT$_d$ when we restrict all its correlators to a linear subspace, or hyperplane, $\mathbb{R}^{p}\subset \mathbb{R}^{d}$ (which is sometimes called ``trivial defect''). By selecting this hyperplane we break the symmetry of the theory from $SO(d+1,1)$ to $SO(p+1,1)\times SO(d-p)$. The group $SO(p+1,1)$ realizes conformal symmetry on $\mathbb{R}^{p}$, while $SO(d-p)$ is understood as a global symmetry under which the CFT$_p$ operators transform.
	The resulting CFT$_p$ however is non-local, meaning that it does not contain a conserved $SO(p)$-spin two operator with dimension $p$ which could be identified as its stress tensor.
	In fact the restriction of the CFT$_d$ stress tensor $T^{\m \n}$ does produce in $p$-dimension a spin-2 operator $T_r^{\a\b}=(T^{\a \b}-\text{trace})$ if we select $\a,\b=1, \dots p$, as well as $SO(p)$ vectors and scalars. However the $T_r^{\a\b}$ does not have the correct scaling dimension, which is still $d$, nor is it conserved. These facts are of course related because conserved spin-2 primary in $d-2$ dimensions should have dimension $d-2$. In fact the stress tensor conservation in $d$ dimension involves derivatives both tangent and orthogonal to $\mathbb{R}^{p}$ and thus does not descend to conservation of $T_r^{\a\b}$.
Physically this is clear: the energy in this setup is not conserved on $\mathbb{R}^{p}$ because it can escape in the orthogonal directions.

	Besides the absence of the stress tensor, the correlation functions of CFT$_p$ are well defined and satisfy the other CFT axioms.
	In particular, the associative OPE of the CFT$_d$ induces an associative OPE in the restricted theory. As a consequence of this, all four point functions of the CFT$_p$ can be expanded in $p$-dimensional conformal blocks (CB$_p$). It follows that any CFT$_d$ correlator, which is expandable in CB$_d$, when restricted can also be expanded in CB$_p$. The way these two facts are compatible is that every CB$_d$ can be expressed as a linear combination of CB$_p$ \cite{Hogervorst:2016hal}. The linear combination is infinite, because an $SO(d+1,1)$ conformal multiplet decomposes into a direct sum of infinitely many $SO(p+1,1)$ conformal multiplets. 
	
	We now come back to our specific problem of Parisi-Sourlas dimensional reduction. This will be understood as a special kind of restriction which will only apply to a subsector of the full Parisi-Sourlas theory. 
	First of all, the restriction will be from the superspace $\mathbb{R}^{d|2}$ to a bosonic $\mathbb{R}^{d-2}$ subspace defined by setting two bosonic coordinates and both fermionic ones to zero:
	\be
	\label{Mcal_dm2}
	\Mcal_{d-2} \equiv \bigg\{y=(x^1,\dots x^{d-2},x^{d-1}, x^{d},\theta,\thetab) \in \mathbb{R}^{d|2} : x^{d-1}=x^{d}=\theta= \thetab=0 \bigg\} \ .
	\ee
	This subspace breaks the symmetry from $OSp(d+1,1|2)$  to  $SO(d-1,1) \times OSp(2|2)$. 
	 
	 Second, dimensional reduction will only apply to the subsector of the supersymmetric theory which consists of $OSp(2|2)$ singlets of the broken symmetry.
Let us illustrate how to construct such operators.
This can be done  by simply considering a superfield $\Ocal(y)$ inserted at a point $y\in\Mcal_{d-2}$ and contracting it with polarization vectors that also live in $\Mcal_{d-2}$ (an analogous construction is detailed in \cite{Billo:2016cpy, Lauria:2018klo} in the context of defect CFTs). For example we may consider a superfield $\Ocal(y,w)$ which transforms in the spin $\ell$ graded symmetric representation of $OSp(d|2)$ (we recall that $w$ is a polarization vector). By restricting both $y$ and $w$ to $\Mcal_{d-2}$ we  obtain an operator $\hat O$ with $SO(p)$-spin $\ell$ :
\be
\label{reductionO}
\left. \Ocal(y,w) \right|_{\Mcal_{d-2}} \to \hat O(\hat x,\hat z)    \, , 
\qquad 
\left. \hat z \equiv w  \right|_{\Mcal_{d-2}}
\, ,
\
\left. \hat x \equiv y  \right|_{\Mcal_{d-2}}
\, .
\ee
We will denote the restricted operators as $\hat O$ and their coordinates as $\hat x\in \mathbb{R}^{d-2}$. 
A hatted operator with indices $\hat O^{\a\b\cdots}$, will automatically imply $\a,\b=1,\dots ,d-2$.
It is clear from the $SO(d-1,1)$ symmetry that the restricted operators belong to a CFT in $d-2$ dimensions. Since $\hat x$ and $\hat z$ are at the origin of the $\mathbb{R}^{2|2}$ space which is fixed by $OSp(2|2)$, the operators defined by \eqref{reductionO} are annihilated by all generators of $OSp(2|2)$, hence they are $OSp(2|2)$-singlets. The definition  \eqref{reductionO} also  implies that if $\mathcal{O}$ is a superprimary with superconformal dimension $\D$ then $\hat O$ transforms under $SO(d-1,1)$ as a CFT$_{d-2}$ primary of dimension $\D$. This follows straightforwardly from \eqref{KDaction}, as $\hat O$ is annihilated by $K^\mu$ $(\mu=1,\cdots,d-2)$ and diagonalizes the $(d-2)$-dilations $D$.

Note that \eqref{reductionO} automatically gives the right symmetry structure of the reduced operators. E.g. a super-traceless graded symmetric $OSp(d|2)$  spin $\ell$ tensor is projected to a traceless symmetric spin $\ell$  tensor in $d-2$ dimensions.  In components, Eq.~\eqref{reductionO} means  that $\hat O$ is obtained from $\Ocal$ by two operations. First, we restrict the $OSp(d+1|2)$ indices to the directions in $\Mcal_{d-2}$. Second, we subtract the traces. The latter operation is implicit in \eqref{reductionO}  because $\hat z^\mu \hat z_{\mu}=0$. We can demonstrate this with the following simple example of a spin $2$ superprimary $\mathcal{O}_{ab}$ which is super-traceless and graded symmetric. We first restrict the indices $a=\mu, b=\nu$ so that $\mu,\nu \in \{1,\ldots , d-2\}$, and then subtract the trace, so that:
\be\label{stresscomp}
\hat O_{\mu\nu}(\hat x) = \mathcal{O}_{\mu\nu}(\hat x) - \frac{g_{\mu\nu}}{d-2}{{\mathcal{O}^\a}_\a }(\hat x) \equiv \mathcal{O}_{\mu\nu}(\hat x) + \frac{g_{\mu\nu}}{d-2}{g_{2|2}^{ab}\mathcal{O}_{ab} }(\hat x) \,,
\ee
Since $\Ocal_{a b}$ is assumed traceless in full superspace, we have equivalently rewriten the subtracted $d-2$ dimensional trace ${\mathcal{O}^\a}_\a$ as minus the trace over the directions orthogonal to $\Mcal_{d-2}$, with $g_{2|2}$ the corresponding $OSp(2|2)$ invariant metric.

The procedure  \eqref{reductionO} works for any operator $\Ocal$ (for a generic mixed symmetric representations we only have to introduce different polarization vectors for each row of their graded Young tableau). 
The restriction to $\theta,\thetab=0$ selects the bottom component $\Ocal_0$ of the superfield as defined in \eqref{Oexpanded}. It follows that the bottom component of any given superprimary operator $\Ocal$ gives rise to an operator $\hat O$ which is a singlet of $OSp(2|2)$. 

We emphasize that construction \eqref{reductionO} can be applied to any operator $\Ocal$ of the Parisi-Sourlas theory. To be sure, there is no requirement that the operator $\Ocal$ should have some protected dimension, or anything of the sort. Thus we obtain a huge number of operators $\hat O$ in $(d-2)$ dimensions. The construction works simplest for scalars when no indices are involved, but also for spinning primaries. 	There is however one subtlety: the attentive reader may have noticed that depending on the dimension $d$, for some operators $\Ocal$ transforming in complicated $OSp(d|2)$ representations, the resulting operator $\hat O$ may come out identically zero. This subtlety does not occur for most commonly occurring representations such as graded symmetric traceless tensors as long as $d\ge 4$ (so that the reduced space has dimension at least 2). We will assume that $d\ge 4$ in the rest of this paper.\footnote{We could in principle also consider the reduction from $3$ to 1 dimensions. Since spin does not exist in 1$d$, the story changes a bit. In this special case scalars superprimaries reduce to scalars, vector superprimaries reduce to parity odd scalars, while any other representation reduces to zero. In particular super stress tensor reduces to zero, so that we do not get a stress tensor in 1$d$. This is of course natural because we do not expect local CFTs in 1$d$. In what follows we will not consider this special case.}

We wish to consider the theory in $(d-2)$ dimensions which consists of primary operators $\hat O$ defined via \eqref{reductionO} and of their $(d-2)$ dimensional descendants. Let us call this space of operators $S_0$. It should be pointed out that there are other $OSp(2|2)$ invariant operators which are not in $S_0$.  E.g.~we can take the trace of operator $\Ocal$ with respect to orthogonal directions (i.e.~the superspace components not in $\Mcal_{d-2}$), or we can take derivatives with respect to these orthogonal directions and contract them with the $OSp(2|2)$ invariant metric $g_{2|2}$.   Let us call operators of these latter types $S_1$. E.g.~in \eqref{stresscomp}, the trace part ${\mathcal{O}^\a}_\a=-g_{2|2}^{ab}\mathcal{O}_{ab}$  and hence can be regarded as an $S_1$ operator. Both $S_0$ and $S_1$ are $OSp(2|2)$ singlets, but it turns out that $S_1$ operators containing $g_{2|2}$ decouple: correlation functions of an arbitrary number of singlet operators with an $S_1$ operator vanish on $\Mcal_{d-2}$. This will be explained in section \ref{sec:decoupling}. Also, exchanges of $S_1$ operators will be shown not to contribute to correlation functions of $S_0$ operators on $\Mcal_{d-2}$. For this reason it makes sense to restrict our attention only to $S_0$ operators.

Now we see that the considered reduction works rather differently from the usually considered trivial defect theories. From each superprimary operator of the $d$-dimensional theory we obtain one primary operator of the $(d-2)$ dimensional theory. In usual trivial defect theories, even if we restrict to the subsector of operators preserved by the defect, reduction is one to many. This is because analogues of operators we called above $S_1$ do not decouple for trivial defects.

Let us now discuss the features of the CFT$_{d-2}$ thus defined by dimensional reduction.
A generic correlation function in the CFT$_{d-2}$ is obtained from that of the $OSp(d+1,1|2)$ theory by restricting to $\mathcal{M}_{d-2}$ as follows: 
	\be
	\label{def_dim_red}
	\la \hat O_1(\hat x_1) \dots  \hat O_n(\hat x_n) \ra  = \la \Ocal_1(y_1) \dots  \Ocal_n(y_n) \ra \bigg|_{\Mcal_{d-2}} \ .
	\ee
In the above equation, $\hat O_i$ is the projection of a superprimary $\mathcal{O}_i$ on $\mathcal{M}_{d-2}$ via \eqref{reductionO}.	
Equation \eqref{def_dim_red} looks like a usual restriction, however the resulting CFT$_{d-2}$ has further special properties. The most surprising one is that it has a conserved stress tensor, as we shall demonstrate in section \ref{stress_tensor_red}. From usual arguments, the reduced theory is endowed with an OPE which arises from the restriction of the full OPE to $\mathbb{R}^{d-2}$. Of course the OPE is closed in the subsector of singlet operators of $OSp(2|2)$, meaning that the OPE of two singlet operators only exchanges singlets (moreover, as we show in section \ref{CB_red}, type $S_1$ operators decouple, so the OPE can be truncated to operators $\hat O$ and their $(d-2)$-dimensional descendants).
	By using the OPE, any four point correlation function of the reduced theory can be expanded in CB$_{d-2}$,  which means there should be a relation between the superblocks of Parisi-Sourlas theory and the blocks of the CFT$_{d-2}$. Quite remarkably, this relation is extremely simple. Indeed, as we prove in section \ref{CB_red}, each superblock is mapped to a single $(d-2)$-dimensional block! 
	So the operators defined as the $OSp(2|2)$-singlet parts of superprimaries  of the Parisi-Sourlas theory give rise to a local CFT in $d-2$ dimensions. 
		
	We have given a general picture of how Parisi-Sourlas dimensional reduction works.
	In the next section we discuss the decoupling of $S_1$ operators. This will be important in the following sections where we prove that the dimensionally reduced theory is indeed local and that the superblocks are equal to conformal blocks in $(d-2)$-dimensions. Additionally, we will show how this equivalence gives a new relation between CB$_{d-2}$ and CB$_d$.
	
\subsection{Decoupling of $S_1$ operators}
\label{sec:decoupling}
	
	As we explained in the last section, there is an important difference between operators $S_0$ and $S_1$, since the latter decouple from the singlet sector. Let us show how this happens.
Recall that $S_0$ consists of primaries generated using \eqref{reductionO} and their $(d-2)$-dimensional descendants. These operators are constructed from the $(d-2)$-dimensional components of $\Ocal$'s and of their $(d-2)$-dimensional derivatives. The objects orthogonal to $\Mcal_{d-2}$ are never involved, thus ensuring the $OSp(2|2)$ invariance of the obtained states. On the other hand, $S_1$ operators are defined by including the orthogonal objects (derivatives or components) but in singlet combinations, i.e. by contracting them with the metric $g_{2|2}$ on the space $\mathbb{R}^{2|2}$ ($d=2$ case of \eqref{metricOSp}). 
Some examples of $S_1$ operators are:\footnote{Recall also the simple example of dimensional reduction \eqref{stresscomp}, where ${\mathcal{O}^\a}_\a= - g_{2|2}^{ab} \Ocal_{a b} $ is an $S_1$ operator.}
\be 
g_{2|2}^{ab} \Ocal_{a b }\, , \qquad
 g_{2|2}^{ab} \partial_a\Ocal_{b} \, ,
\qquad 
g_{2|2}^{ab}\partial_a  \partial_b \Ocal  \, .
\ee
With these definitions, $S_0$ and $S_1$ operators span the space of all $OSp(2|2)$ singlets.

Now we will explain that, within singlets, the $S_1$ operators completely decouple on $\Mcal_{d-2}$ in the following sense. Consider a correlation function of an arbitrary $S_1$ operator and any number of singlets (which may be $S_0$ or $S_1$). We claim that this correlator vanishes on $\Mcal_{d-2}$:
\be
\label{decoupling}
\langle \tilde \Ocal \,  \Ocal_1 \dots  \Ocal_n \rangle = 0 \,\text{ on } \Mcal_{d-2}, \qquad \mbox{for $\tilde \Ocal\in S_1$, $ \Ocal_i\in S_0 \oplus S_1$. }
\ee
We repeat that this is only true if all operators are singlets and they all are positioned on $\Mcal_{d-2}$. Because of this limitations, it would be wrong to say that the $S_1$ operators are identically zero, although they do decouple in the described sense. 

Vanishing of \eqref{decoupling} is proven as follows. By definition of $S_1$ operators, \eqref{decoupling} is some $OSp(2|2)$-invariant expression containing at least one metric $g_{2|2}$, and which depends on coordinates of operators, all in $\Mcal_{d-2}$. When constructing the invariant, the indices of $g_{2|2}$ may be contracted with the reduced space objects (e.g. $\hat x,\hat z$ or the reduced metric), which will trivially give zero. Alternatively, the indices of $g_{2|2}$ may be contracted with the indices of some other $g_{2|2}$ floating around, such as if there are more than one $S_1$ operators in \eqref{decoupling}. However, the latter contraction gives the supertrace $(g_{2|2})_a^{\phantom{a} a}=2-2=0$, which also vanishes. 
In other words it is not possible to build a nontrivial $OSp(2|2)$ singlet out of $g_{2|2}$ and reduced tensors. Hence all singlet correlation functions involving an operator $S_1$ must vanish.

The lack of singlets built out of $g_{2|2}$ plays a crucial role in all proofs related to dimensional reduction.
E.g.~this fact is hidden in various steps of the demonstrations of locality (see section~\ref{stress_tensor_red}) and dimensional reduction of the superblocks (see section~\ref{CB_red}).

\subsection{The stress tensor multiplet}
\label{stress_tensor_red}
In this section we shall demonstrate that, assuming that the $OSp(d+1,1|2)$ theory is local,  the singlet sector of its restriction to $\mathcal{M}_{d-2}$ is also local.
The first step is to show that if the super-conformal theory has a super stress tensor $\mathcal{T}_{ab}$, then the dimensionally reduced theory also has a spin two operator with dimension $d-2$, which satisfies a conservation equation.

We discussed in section \ref{subsec:symCFT} the super stress tensor $\mathcal{T}_{ab}$ and its conservation. 
 Following the rule \eqref{reductionO}, the dimensionally reduced stress tensor  $\hat T$ will be related to $\mathcal{T}$ by the following equation:
\be
\label{lowerstress}
\hat T(\hat x, \hat z) \equiv \left. \mathcal{T}(y,w)\right|_{\Mcal_{d-2}}  \,.
\ee
Setting the transverse coordinates, in particular $\theta,\bar\theta$, to zero, picks out the bottom component $\mathcal{T}_0$ which has the same dimension as the $\mathcal{T}_{ab}$ superdimension, that is $d-2$. 
Since $\hat T$ has the correct conformal dimension and spin, it is a good candidate for the stress tensor of the CFT$_{d-2}$. 
Let us see that the conservation of $\mathcal{T}$ implies conservation of $\hat T$. We have
\be
\label{eq:curlyTcons}
0 = \partial^a\mathcal{T}_{a\nu}=\sum_{\a=1}^{d-2} \partial_\a g_{d-2}^{\a\b}\mathcal{T}_{\b \nu}+\partial_a g_{2|2}^{ab}\mathcal{T}_{b\nu} \, ,
 \ee
where we denoted by $g_{d-2}$ and $g_{2|2}$ the metrics along and orthogonal to $\mathcal{M}_{d-2}$. The second term is an operator of type $S_1$ in the sense of the previous section. In the first term we have $\mathcal{T}_{\b \nu}=\hat{T}_{\b \nu}$ modulo another $S_1$ operator, see \eqref{stresscomp}.
So we conclude that $\partial^\a \hat{T}_{\a\b}=0$ modulo $S_1$ operators. 
As we showed in section \ref{sec:decoupling}, correlation functions of $S_1$ with other singlets vanish on $\mathcal{M}_{d-2}$. 
So $\hat T$ is conserved within the singlet sector.\footnote{The just given argument can be succinctly expressed using the differential operator \eqref{Tod}, which ``opens up'' indices of tensors contracted with the external polarization vectors, by showing that $(\partial_y \cdot D_w) \mathcal{T}(y,w) = 0$ implies $ (\partial_{\hat x} \cdot D_{\hat z})\hat T(\hat x, \hat z) = 0$ in the singlet sector (where $D^{\m}_{\hat z}$ is used to open the indices of $SO(d-2)$ tensors). In fact, one can see that on $\Mcal_{d-2}$, the operator $\partial_y \cdot D_w$ acts as $\partial_{\hat x} \cdot D_{\hat z}$ 
 modulo terms of $S_1$-type.} 

So far we only discussed conservation of $\hat{T}_{\mu\nu}$ away from other insertions.
In the following we would like to see how the contact terms dimensionally reduce.
This would fully prove that the stress tensor candidate $\hat T_{\mu\nu}$ generates translations in the CFT$_{d-2}$, hence it is the true stress tensor, and the CFT$_{d-2}$ is local. 
In  \eqref{Tlocal}  we defined the meaning of locality for a Parisi-Sourlas theory. 
In what follows we use an alternative definition of locality which repackages equation  \eqref{Tlocal}  in terms of a condition on the OPE between $\mathcal{T}_{ab}$ and $\Ocal$. This is done by analogy with the non-supersymmetric story in \cite{Cardy:1987dg} (see also the lucid discussion of such matters in \cite{Simmons-Duffin:2016gjk}).
Indeed \eqref{Tlocal} is related to the following OPE contribution:
\be\label{Tope}
\mathcal{T}_{ab}(y) \Ocal(0)\sim \cdots + {B_{ab}}^c(y)\partial_c\Ocal(0)+\cdots\,,
\ee
where the tensor structure $B$ has to satisfy the following property:
\be\label{Tprop}
\int_ V  \dd^dx  \dd \thetab \dd \theta \ \partial^a B_{abc}(y)=g_{cb}\,,
\ee
where $ V $ is the interior of a supersphere with radius $\varrho$. 

We would like to show that, if \eqref{Tope} and \eqref{Tprop} hold, then the operator ${\hat T}_{\mu\nu}$  has the following OPE with $\hat O$:
\be\label{Topelow}
{\hat T}_{\mu\nu}(\hat x) \hat O(0)\sim \cdots + {B_{\mu\nu}}^\rho(\hat x)\partial_\rho \hat O(0)+\cdots\,,
\ee
with the condition
\be\label{Tproplow}
\int_{ \hat V} \dd^{d-2}\hat x \ \partial^\mu B_{\mu\nu\rho}(\hat x)=g_{\nu\rho}\, ,
\ee
where $ \hat V$ is the projection of $ V $ on $\mathcal{M}_{d-2}$\,. These equations then imply that $\hat T$ generates $(d-2)$ dimensional translations.

Before entering the proof, it is convenient to fix the structure of $B_{abc}(y)$. This is determined by requiring ${B^a}_{ac}=0$ and $\partial^a B_{abc}=0$ (away from $y=0$), respectively from the tracelessness and conservation of the super stress tensor, as well as scaling as $~1/y^{d-2}$ for dimensional reasons. This implies:
\be\label{Bstruct}
B_{abc}(y)=C \Big [\frac{y_a g_{cb}+y_b g_{ac} -\frac{2}{d-2}y_c g_{ba}}{{(y^ay_a)}^{\frac{d}{2}-1}} +(d-4) \frac{(y_a y_b-\frac{g_{ba}}{d-2}y^ey_e)y_c}{{(y^ay_a)}^{\frac{d}{2}}}\Big]\,,
\ee
with an overall normalization constant $C=\frac{d-2}{2(d-3) S_{d-2}} $ (where $S_{d-2}$ is the area of the sphere in $d-2$ dimensions), which we computed by performing the integral \eqref{Tprop} explicitly. 

Now that $B$ is fixed, it is easy to show that the OPE \eqref{Topelow} follows directly from \eqref{Tope}, by simply projecting it to $\mathcal{M}_{d-2}$.
Indeed $a,b$ are trivially projected to $\mu,\nu$ and it is easy to see that ${B_{\mu\nu}}^c \partial_c \Ocal= {B_{\mu\nu}}^\rho  \partial_\rho \Ocal$ (in fact ${B_{\mu\nu}}^c$ vanishes when $y=\hat x\in \mathcal{M}_{d-2}$ and $c$ is an index orthogonal to $\mathcal{M}_{d-2}$). 

We are then left to show how \eqref{Tproplow} follows from \eqref{Tprop}.
For this it is convenient to use the following equation:
\be\label{Radial2}
\int  \dd^{d}x   \dd \thetab   \dd \theta f(w_i^\mu x_\mu, y^2) = \int  \dd^{d-2}\hat x   f(w_i^\mu \hat x_\mu, {\hat x}^2)\,,
\ee
where $y^2=x^2+2\thetab\theta$ and $w_i^\mu$ are vectors in $\mathcal{M}_{d-2}$.
To derive it we apply the $d=2$ case of \eqref{RadialInt1} integrating over $x^{d-1},x^{d},\theta,\thetab$ and treating the rest of the variables as spectators (since $w_i^\mu\in \mathcal{M}_{d-2}$, the function being integrated is $OSp(2|2)$ invariant). 

Eq.~\eqref{Tprop} can be put in the form \eqref{Radial2} as follows. 
First we note that in \eqref{Radial2} the integral is over the whole $\mathbb{R}^{d|2}$ while in \eqref{Tprop} it is over the super-ball $V$. We just replace the latter with an integral over $\mathbb{R}^{d|2}$ with an appropriate step-function $\Theta(\varrho- \sqrt {y^2})$.
Then we contract the indices of \eqref{Tprop} with polarization vectors $w_1^b,w_2^c$ that are restricted to $\mathcal{M}_{d-2}$.
In summary we use \eqref{Radial2} with $f \equiv \Theta(\varrho-\sqrt {y^2})\partial^a B_{a b c} w_1^b w_2^c $. 
Finally we need to show that $\partial^a B_{a\nu\rho}(y)|_{y\in \mathcal{M}_{d-2}}=\partial^\mu B_{\mu\nu\rho}(\hat x)$. This is easy to see from a direct computation and basically descends from the fact that the superdivergence $\partial^a y_a=d-2$ is equal to the divergence $\partial^\mu \hat{x}_\mu$ in $\mathbb{R}^{d-2}$.
Putting together all the ingredients, the equation \eqref{Tprop} reduces to
\be\label{Tobs}
g_{\nu\rho}=\int_ V  \dd^dx  \dd \thetab \dd\theta  \ \partial^a B_{a\nu\rho}(y)=\int_{ \hat V} \dd^{d-2}\hat x \ \partial^\mu B_{\mu\nu\rho}(\hat x)\, ,
\ee
which proves the condition \eqref{Tproplow}. Not surprisingly, the dimensionally reduced $B$ has the same functional form as that one given by Cardy \cite{Cardy:1987dg}.

The proof above may look technical, but it is rather explicit and has the advantage of clearly showing how the contact terms are dimensionally reduced. We thus reach the important conclusion that the locality of the Parisi-Sourlas CFT implies the locality of the CFT$_{d-2}$. Here it is important to stress that by CFT$_{d-2}$ we are referring to the $OSp(2|2)$-singlet sector of the restricted theory. Indeed for operators which are charged under $OSp(2|2)$ we would not be able to use \eqref{Radial2}, and our proof would not go through.

	\subsection{OPE and conformal blocks}
	\label{CB_red}
	
	\subsubsection{Dimensional reduction of OPE}
	\label{OPE_red}

 Let us show the implications of the decoupling \eqref{decoupling} for the OPE. 
Consider the OPE of two scalar primaries $\hat O_i \in S_0$ of dimensions $\D_i$. 
We can obtain this OPE by dimensionally reducing the superspace OPE \eqref{super_OPE}, which gives the expression of the form:
\be
\label{OPEparallel11}
\hat O_1(\hat x)  \hat O_2(0) =  \sum_{\Ocal}  \frac{\l_{12 \hat \Ocal}}{|\hat x|^{\D_1+\D_2-\D+\ell}} [\hat x_{\a_\ell} \cdots \hat x_{\a_1}  \Ocal^{\a_1\ldots \a_\ell}(0) + \textrm{superdesc.}]
\, .
\ee
We can further rearrange the r.h.s~of this equation. First of all we want to replace the components of $\Ocal$ by the components of the $d-2$ dimensional traceless symmetric primary $\hat O$ defined via \eqref{reductionO}. As explained in section \ref{dimredsec}, this involves some subtracted traces which are $S_1$ operators. We also want to split the superdescendant contributions into those of usual $d-2$ dimensional descendants and the rest, i.e.~the superdescendants involving superderivatives orthogonal to $\Mcal_{d-2}$. The latter are also $S_1$ operators: they must appear in $OSp(2|2)$ singlet combinations, hence will involve contracting superderivatives with the $g_{2|2}$ metric. Thus after rearrangement we get:
	\be
\label{OPEparallel1}
\hat O_1(\hat x)  \hat O_2(0)=  \sum_{\hat O\in S_0}  \frac{\l_{12\hat O}}{|\hat x|^{\D_1+\D_2-\D+\ell}} \ [\hat x_{\a_1} \cdots \hat x_{\a_\ell} \hat O^{\a_1 \dots \a_\ell}(0) + \textrm{desc.}] 
+ \text{$S_1$ contribution}
\, ,
\ee
So, any superprimary $\Ocal$ with dimension $\D$ and $OSp(d|2)$-spin $\ell$ in the superOPE \eqref{super_OPE} gave rise to an $S_0$ primary $\hat O$ contributing to the OPE \eqref{OPEparallel1}, of the same scaling dimension and with the same OPE coefficient: $\l_{12 \hat O}=\l_{12 \Ocal}$. This primary is accompanied by its $(d-2)$-dimensional descendants, as indicated in \eqref{OPEparallel1}. 

Now suppose we are interested in correlation functions of $\hat O_i \in S_0$, e.g.~their four point function. To compute the four point function, we use OPE \eqref{OPEparallel1} twice. Then we are reduced to two point functions of exchanged operators. However as noticed above the $S_1$ operators have vanishing two point functions with any singlet, in particular among themselves, and will drop out. Thus, for purposes of any such computation we can truncate the OPE \eqref{OPEparallel1} dropping the $S_1$ contribution:
	\be
	\label{OPEparallel}
	\hat O_1(\hat x)  \hat O_2(0) =  \sum_{\hat O}  \frac{\l_{12\hat O}}{|\hat x|^{\D_1+\D_2-\D+\ell}} \ [\hat x_{\a_1} \cdots \hat x_{\a_\ell} \hat O^{\a_1 \dots \a_\ell}(0) + \textrm{desc.}] \ ,
	\ee
It is in this sense that the OPE of $S_0$ operators is closed on themselves, as already anticipated in section \ref{dimredsec}.
	
It is nice to rephrase this conclusion in terms of conformal blocks. Indeed, since the $S_1$ operators decouple, the superblocks will have a very simple relation with the CB$_{d-2}$. This will be discussed in the next subsection.

	\subsubsection{Dimensional reductions of superconformal blocks}
	In this section we want to prove that the $OSp(d+1,1|2)$ conformal blocks are equal to conformal blocks in $d-2$ dimensions. Our strategy is to show that the two functions satisfy the same differential equation, which arises by applying the (super) conformal Casimir. 
	In order to make the computation easier the argument is formulated in super-embedding space, where the the super-Casimir operator takes a very simple form.
	
	For simplicity we will first focus on a four point function of scalar superprimaries and we will analyze the super-Casimir equation \eqref{casPDE} for the exchange of an operator $\Ocal$ with super-dimension $\D$ and which transforms in a graded symmetric representation of $OSp(d|2)$ with spin $\ell$:
	\be
	C \; G^{d|2}_{\D,\ell}(P_i)=c^{d|2}_{\D,\ell}\ G^{d|2}_{\D,\ell}(P_i)\ ,
	\ee
	where $P_i$ are points defined in super-embedding space of section \ref{subsec:SuperEmbeddingCorrelationFunctions}. The eigenvalue appearing in this equation is given in Eq.~\eqref{EigenvalueCasimir} and, as mentioned there, is equal to the eigenvalue of the $d-2$ dimensional Casimir equation. We are thus left to check that action of the differential operator $C$ takes the same form as the Casimir differential operator in $d-2$ dimensions when we restrict this equation to the submanifold $\Mcal_{d-2} $ defined in section \ref{Mcal_dm2}.
 This must be true, since the Casimir differential operators are equal up to terms belonging to $S_1$, which vanish according to \eqref{decoupling}. 
However, to be more transparent, let us show in more details how this match takes place.
 
	For the purpose of the proof, it is convenient to write the superconformal Casimir $C$, see Eqs.~\eqref{Casimir_Op}, \eqref{def:LAB}, as follows:
	\be
	\label{sum_C}
	C= C_{P_1}+C_{P_2}+C_{P_1,P_2} \ ,
	\ee
	where the operators $C_{P}$ and $C_{P,Q}$ are defined as
	\be
	\label{CPiCP1P2}
	C_{P}\equiv (d-2+P \cdot \partial_P)P \cdot \partial_{P} - \underbrace{P^2 \partial_{P} ^2}_{=0}
	\ ,
	\qquad 
	C_{P,Q} \equiv 2 P^{A} Q^{B} (\partial_{P})_{B} (\partial_{Q})_{A} - 2 (P \cdot Q)(\partial_{P}\cdot \partial_{Q}) \ ,
	\ee
	where the term $P^2 \partial_{P}^2$ can be dropped as shown since $P^2=0$ on the projective lightcone. The terms in \eqref{sum_C} arise  by collecting the contributions $(L_1)^2$, $(L_2)^2$ and $ L_1 L_2$ in \eqref{Casimir_Op} and by commuting all the derivatives to the right.\footnote{
		To this end it is useful to apply the following formula
		\be
		\frac{1}{2} \bar g^{AD}\bar g^{BC} L(P_i)_{AB}L(P_j)_{CD}=P_i^{A}\partial_i^{B} (P_j)_{B} (\partial_j)_{A}-(-1)^{[C] [D]} \bar g^{AC}\bar g^{BD} (P_i)_{A} (\partial_i)_{B} (P_j)_{C} (\partial_j)_{D} \ .
		\ee
		Other useful relations are collected in appendix \ref{app: superspace}.
	}
	The relations \eqref{CPiCP1P2} are also useful to explicitly check the expression for the eigenvalue \eqref{EigenvalueCasimir}. In fact the action of Casimir on $\Ocal(P,Z)$ takes the form $C_{P}+C_{Z}+C_{P,Z}$, where $C_{P,Z}$ reduces to $2 Z\cdot \partial_Z$. By using the scaling properties of $\Ocal(P,Z)$ one directly recovers \eqref{EigenvalueCasimir}.
	
	We now want to consider the action of $C$ on the function $G^{d|2}_{\D,\ell}(P_i)$ when we restrict the points $P_i$ on the manifold $\Mcal_{d-2}$.
	With this restriction, we can simply drop all the terms which give a contribution proportional to  $P^{d-1},P^{d}, \theta, \thetab$. 
	Then it is easy to see that the operators $C_{P_i}$ reduce to
	\be
	C_{P_i} G^{d|2}_{\D,\ell}(P_j)\big |_{
		\Mcal_{d-2}
	}= \left(d-2+ \sum_{\alpha=0}^{d-2} P^\alpha_i \cdot \partial_{P^\alpha_i}\right) \sum_{\alpha=0}^{d-2} P^\alpha_i  \partial_{P^\alpha_i}  \  G^{d|2}_{\D,\ell}(P_k)\big |_{
		\Mcal_{d-2}} \ ,
	\ee
	which takes the same form as their $d-2$ dimensional counterpart. For the operator $C_{P_1,P_2}$ the proof is slightly more complicated.
	By dropping terms  proportional to $x_{d-1},  x_{d}, \theta, \thetab$ we only obtain
	\begin{align}
	C_{P_1,P_2} G^{d|2}_{\D,\ell}(P_j)\big |_{
		\Mcal_{d-2}} =
	&-  \left( 2
	\sum_{\alpha=0}^{d-2}  P^\alpha_1 \cdot P_{2 \, \alpha} 
	\sum_{\alpha=0}^{d-2} 
	\partial_{P^\alpha_1} \partial_{P_{2 \alpha}}  -2 \sum_{\alpha,\beta=0}^{d-2} P_1^{\alpha} P_2^{\beta} (\partial_{P_1})_{\beta} (\partial_{P_2})_{\alpha} \right)  G^{d|2}_{\D,\ell}(P_j)\big |_{ 
		\Mcal_{d-2}} 
	\nonumber
	\\
	&-  2 
	\sum_{\alpha=0}^{d-2}  P^\alpha_1 \cdot P_{2 \, \alpha}
	\left[  
	(\partial_{P^{d-1}_1} \partial_{P^{d-1}_{2}}+\partial_{P^{d}_1} \partial_{P^{d}_{2}}) - ( \partial_{\theta_1}\partial_{\bar \theta_2}+\partial_{\theta_2} \partial_{\bar \theta_1}) \right]  G^{d|2}_{\D,\ell}(P_j)\big |_{ 
		\Mcal_{d-2}} \ .
	\nonumber
	\end{align}
	The first line reproduces exactly the term needed for dimensional reduction to hold. 
	It is straightforward to show that the second line vanishes.
	Indeed, by conformal invariance, the blocks are functions of the scalar products $P_i \cdot P_j$.
	The two terms in the square brackets have the same action on any function $f(P_i \cdot P_j)\big |_{\Mcal_{d-2}}$, with opposite signs, so that their contribution exactly cancels.\footnote{To make the argument more explicit let us consider some  examples. 
		First we act with the square bracket on a term $(P_1 \cdot P_3)(P_2 \cdot P_4)$. The result of the action is $P^{d-1}_{3} P^{d-1}_{4}+P^{d}_{3} P^{d}_{4}-\theta_3\thetab_4-\theta_4\thetab_3 $, which vanishes on $\Mcal_{d-2}$ because all terms are projected to zero. 
		In order to obtain a result that is not trivially projected to zero, one can consider the action of the square bracket on $(P_1 \cdot P_2)$. However this would give a term proportional to the super-trace of the orthogonal space which is zero, $(1+1)-(1+1)=0$. It is easy to see that the action of the square bracket on any function of $(P_i \cdot P_j)$ will either be proportional to terms which vanish on  $\Mcal_{d-2}$ or give the vanishing super-trace result.
		 Indeed this is only an explicit example of the decoupling described in section~\ref{sec:decoupling}.
	}

	We therefore proved that the super-Casimir differential equation restricted to $\Mcal_{d-2}$ is equal to the Casimir equation in $d-2$ dimensions. It is also straightforward to show that the boundary condition at short distances that the conformal partial waves have to satisfy, and which follows from the OPE, is the same.
	Hence, the conformal partial waves $ G^{d|2}_{\D,\ell}(P_i)$ are identical to the partial waves $ G^{d-2}_{\D,\ell}(\hat P_i)$ of a $d-2$ theory when the points $P_i=\hat P_i$ are restricted to $\Mcal_{d-2}$. This implies that conformal blocks, to which conformal partial waves are proportional as shown in Eq.~\eqref{casPDE}, should agree as functions of $u,v$:
	\be
	\label{g_reduction}
	g^{d|2}_{\Ocal}(u,v) =g^{d-2}_{O}(u,v) \ .
	\ee
Indeed, by conformal invariance, knowing conformal blocks on $\Mcal_{d-2}$ is enough to fix their functional form completely. Note that in the main case of our interest $d\ge4$, $d-2\ge 2$, we have two independent cross ratios $u,v$ both before and after reduction.

	The only difference between the blocks is the meaning of the argument at which they are usually evaluated: $g^{d-2}$ at the standard cross ratios $u,v$, the superblock $g^{d|2}$  at the super-cross-ratios $U,V$ of \eqref{supercrossratios}.

	So far we only considered the case of a spin $\ell$ exchange in a scalar four point function, however we can easily run the same argument for generic spinning external operators and for the exchange of an operator in a generic $OSp(d|2)$ representation.
	For example, given a four point function of operators $\Ocal_i(P_i,Z_i)$ with spin $\ell_i$, the super-Casimir operator in the embedding space can be written as a sum of terms $C_{P_i}$, $C_{Z_i}$, $C_{Z_i,Z_j}$, $C_{P_i,Z_j}$, $C_{P_i,P_j}$ as defined in  \eqref{CPiCP1P2}. For all such terms we already proved  that dimensional reduction works as long as they act on scalar functions of $P_i$ and $Z_j$ (by superconformal invariance the superblocks are scalar functions dependent only on the combinations $P_i\cdot P_j$, $P_i\cdot Z_j$, $Z_i\cdot Z_j$) and as long as we set $P_i,Z_j \in \Mcal_{d-2}$.
	
	With this simple argument we thus proved a very general result: the equality of superconformal partial wave and usual CFT$_{d-2}$ partial waves. There are some exceptions to the above rule.
E.g.~when the exchanged operator belongs to a representation which does not exist in $d-2$ dimensions. In this case the associated conformal partial wave vanishes when restricted to $\Mcal_{d-2}$.\footnote{\label{note:dual}Notice however that some representations which naively do not exist in $d-2$ dimensions, may be dualized to allowed representation, and then this vanishing does not occur. E.g.~the representation with $\ell$ Young tableau boxes in the first row and 1 box in the second row naively is not realized in three dimensions, however by contracting it to an epsilon tensor we can transform it to a Young tableau which exists; it corresponds to a parity odd traceless and symmetric representation of spin $\ell$.}
	See also the discussion in section \ref{sec:comments}.
	Another case is when the kinematics of the reduced space does not allow for the complete reconstruction of the superconformal partial wave. 
	For example by going from 3d to 1d we should interpret \eqref{g_reduction} as a relation between the superblock restricted to a line and the 1d block (where both of them depend on a single cross ratio).

	Equality of conformal blocks for common representations has very important implications for the understanding of dimensional reduction, which we will discuss in section \ref{sec:comments}. Before doing so, we show how this result also has interesting consequences for the theory of conformal blocks.
	\subsubsection{Relations between conformal blocks in different dimensions}
	In the previous section we showed that the  $OSp(d+1,1|2)$ superconformal blocks are equal to blocks in $d-2$ dimensions. 
	On the other hand, it is a completely standard fact that the superconformal block can be decomposed as a sum of regular $d$-dimensional blocks which sum up the contributions of all the conformal primaries in the supermultiplet of the exchanged superprimary operator. Combining these two facts we are led to a somewhat unexpected conjecture: there should be a linear relation expressing a single conformal block in $d-2$ dimensions as a \emph{finite} linear combination of conformal blocks in $d$ dimensions. Normally, relations between conformal blocks are studied the other way around. Indeed, reduction of a $d$ dimensional CFT to a trivial $d-2$ dimensional defect implies that a single $d$ dimensional block should be expressible as an \emph{infinite} linear combination of $d-2$ dimensional blocks. Such infinite linear combinations have been worked out by Hogervorst \cite{Hogervorst:2016hal} for reduction $d\to d-1$, and by using his formulas twice we will get a doubly infinite sum for reduction $d\to d-2$. We thus claim that these doubly infinite sums can be inverted by a finite linear combination!
	
	Surprising as it is, such a magic finite relation indeed exists, and it takes the following form, beautiful to the eye of any conformal block expert:
	\be
	g^{(d-2)}_{\D , \ell}= g^{(d)}_{\D , \ell}+   c_{2,0}\, g^{(d)}_{\D+2 , \ell} + c_{1,-1}\, g^{(d)}_{\D+1 , \ell-1} + c_{0,-2} \, g^{(d)}_{\D , \ell-2} + c_{2,-2} \, g^{(d)}_{\D+2 , \ell-2} \ ,
	\label{gd-2togd}
	\ee
	where the generic scalar block in $d-2$ dimensions is written as a linear combination of only five  blocks in $d$ dimensions. 
	The coefficients can be written in closed form as follows:
	\begin{align}
		c_{2,0}&=\textstyle -\frac{(\Delta -1) \Delta  \left(\Delta -\Delta _{12}+\ell\right) \left(\Delta +\Delta _{12}+\ell\right) \left(\Delta -\Delta _{34}+\ell\right)
			\left(\Delta +\Delta _{34}+\ell\right)}{4 (d-2 \Delta -4) (d-2 \Delta -2) (\Delta +\ell-1) (\Delta +\ell)^2 (\Delta +\ell+1)} \, ,\nonumber \\[5pt]
		c_{1,-1}&=\textstyle-\frac{(\Delta -1) \Delta _{12} \Delta _{34} \ell}{(\Delta +\ell-2) (\Delta +\ell) (d-\Delta +\ell-4) (d-\Delta +\ell-2)} \, ,	\label{coefficientsCBdirred}\\[5pt]
		c_{0,-2}&=\textstyle-\frac{(\ell-1) \ell}{(d+2 \ell-6) (d+2 \ell-4)} \, ,\nonumber\\[5pt]
		c_{2,-2}&=\textstyle
		\frac{(\Delta -1) \Delta  (\ell-1) \ell \left(d-\Delta -\Delta _{12}+\ell-4\right) \left(d-\Delta +\Delta _{12}+\ell-4\right) \left(d-\Delta
			-\Delta _{34}+\ell-4\right) \left(d-\Delta +\Delta _{34}+\ell-4\right)}{4 (d-2 \Delta -4) (d-2 \Delta -2) (d+2 \ell-6) (d+2 \ell-4) (d-\Delta +\ell-5)
			(d-\Delta +\ell-4)^2 (d-\Delta +\ell-3)} \, .\nonumber
	\end{align}
	where as usual $\Delta_{ij}=\Delta_i-\Delta_j$ are dimension differences of the external scalar primaries.
	This assumes conformal blocks normalized as in \cite{Dolan:2000ut, Dolan:2003hv,Penedones:2015aga} (see Table I in \cite{Poland:2018epd} for a comparison of different normalizations).
We were able to obtain these relations in generic dimensions even though the conformal blocks are not known in a closed form.
	To do so we used the recurrence relation \cite{Kos:2013tga, Penedones:2015aga} which determines the conformal blocks as an expansion in the radial cross ratio $r$ of \cite{Hogervorst:2013sma}.
	In fact all the coefficients $c_{i,j}$ (for arbitrary values of the spin $\ell$) of the ansatz  \eqref{gd-2togd} are completely fixed  already at order $O(r^2)$.
	We then checked that higher orders in $r$ also agree with Eqs. (\ref{gd-2togd}), (\ref{coefficientsCBdirred}).\footnote{We also checked that Eqs. (\ref{gd-2togd}) and (\ref{coefficientsCBdirred}) are satisfied by some closed form expressions of conformal blocks. In particular for the relation between $2$ and $4$ dimensions---where both blocks are known from the early work of \cite{Dolan:2000ut, Dolan:2003hv}---and for $1$ and $3$ dimensions. In the latter case the $1d$ blocks depend on a single cross ratio and should be compared to the $3d$ diagonal blocks (for $\ell=0$) of \cite{ElShowk:2012ht, Hogervorst:2013kva}.}
	
Our relation (\ref{gd-2togd}) is reminiscent of a relation between conformal blocks in different dimensions found by Dolan and Osborn (see Eq. (5.4) of \cite{Dolan:2003hv} and Eq.~(4.42) of \cite{Dolan:2011dv}). However the two relations are different. In fact the latter relates one block in $d$ dimensions to five blocks in $d-2$ dimensions (so it works in the opposite way). Moreover the $d$-dimensional block is multiplied by some function of the cross ratios, while in our relation all coefficients are pure numbers. Although the connection is not obvious, it turns out that by judiciously combining Eq. (5.4) of \cite{Dolan:2003hv} with Eqs.~(5.1), (5.2) of the same article, one can eliminate the prefactor and obtain our recursion relation.\footnote{We thank Hugh Osborn for pointing this out and for sending us the derivation after our paper appeared \cite{Hugh}.}  
	
	Eq.~\eqref{gd-2togd} is very useful to demonstrate some features of the $OSp(d+1,1|2)$ representation theory, which we explained in the previous sections.
	For example, we may look at the expansion of a superfield in the Grassmann variables, as shown in \eqref{Oexpanded}. An operator $\Ocal^{a_1\dots a_\ell}$ in traceless graded-symmetric representation of $OSp(d|2)$ of spin $\ell$, contains exactly five classes of Grassmann-even primary operators (which transforms under $SO(d)$):
	\be
	\Ocal^{a_1\dots a_\ell} \supset 
	\underbrace{\Ocal_0^{\a_1\dots \a_\ell}}_{\D, \ell}
	\,, \quad
	\underbrace{ \Ocal_{\theta \thetab}^{\a_1\dots \a_\ell} }_{\D+2, \ell}
	\,, \quad
	\underbrace{ 
		\Ocal_{\theta}^{\theta \a_2 \dots \a_\ell}, 
		\Ocal_{\theta}^{\thetab \a_2 \dots \a_\ell} 
		\Ocal_{\thetab}^{\theta \a_2 \dots \a_\ell} 
		\Ocal_{\thetab}^{\thetab \a_2 \dots \a_\ell} 
	}_{\D+1, \ell-1}
	\,, \quad
	\underbrace{ \Ocal_{0}^{\theta \thetab \a_3 \dots \a_\ell} }_{\D, \ell-2}
	\,, \quad
	\underbrace{ \Ocal_{\theta \thetab}^{\theta \thetab \a_3 \dots \a_\ell} }_{\D+2, \ell-2} \ .
	\ee
	Since the tensor indices $\theta$ and $\thetab$ can at most appear once (because of graded-symmetry), primaries with $\ell-3,\ell-4, \dots$ are forbidden.  
	Hence, from this simple analysis, we obtain the form of \eqref{gd-2togd}. Of course for low spins $\ell=0,1$, some of the five mentioned classes of representations vanish. For these low spins the corresponding coefficients $c_{i,j}$ vanish, as can be seen from \eqref{coefficientsCBdirred}.
	
	Another interesting feature of the coefficients $c_{i,j}$ is that they are not sign-definite, in fact they are all negative for large $\Delta$. This is related to the fact that the Parisi-Sourlas theory is non-unitary. Notice that the above-mentioned infinite reductions from larger to smaller $d$ give rise to expressions with positive coefficients.
	
	We can perform an even more refined consistency check, by obtaining the exact form of the coefficients \eqref{coefficientsCBdirred} by studying the  decomposition of the superconformal multiplets into conformal multiplets. As an example we recover the coefficient $ c_{2,0}$ for the case of $\ell=0$, which multiplies $g^{(d)}_{\D+2 , \ell=0}$. 
	This conformal block is associated to a scalar primary operator $\tilde\Ocal$ which is a descendant at the level two of a scalar superprimary $\Ocal$. In order to have the right quantum numbers the operator $\tilde\Ocal$ has to have the form:
	\be
	|\tilde\Ocal \rangle = (P^\a P_\a + a \ P^{\theta}P^{\thetab}) |\Ocal \rangle \ ,
	\ee
	where the coefficient $a$ can be fixed by requiring that $\tilde\Ocal$ is a primary operator, $K^\m |\tilde\Ocal \rangle = 0$. We thus find 
	\be
	a =2\D -d +2 \ .
	\ee
	The coefficient $c_{2,0}$ arises because of the different normalization of $\tilde\Ocal$, which is a superdescendant, and therefore is not canonically normalized as a primary operators.
	In particular the coefficient $c_{2,0}$ is written as 
	\be
	\label{c20FromMultDecomp}
	c_{2,0}(\ell=0)= \frac{\OPE(\D_{12}) \OPE(\D_{34})}{\NORM} \ ,
	\ee
	where the coefficients $\NORM$ and $\OPE$ are respectively defined in terms of the norm of $|\tilde\Ocal \rangle$ and its three point functions with other two operators:
	\be
	\langle \tilde\Ocal|\tilde\Ocal \rangle \equiv \NORM \ \langle \Ocal|\Ocal \rangle \ ,
	\qquad
	\langle  \Ocal_i \Ocal_j \tilde\Ocal \rangle \equiv \OPE(\D_{ij}) \ \langle  \Ocal_i \Ocal_j \Ocal \rangle \ .
	\ee
	By using the commutation relations of the superalgebra \eqref{susyalgebra} (or equivalently by taking derivatives on a two point function), we obtain
	\be
	\label{NORMcoeff}
	\NORM=-4 \Delta  (\Delta +1) (d-2 (\Delta +1)) (d-2 (\Delta +2)) \ .
	\ee
	In order to compute $\OPE$ we conveniently choose to act with the combination $ (P^\a P_\a + a \ P^{\theta}P^{\thetab}) $ on the leading OPE $\Ocal(x,\theta,\thetab)\Ocal_i(0) \sim c_{\Ocal i j} (x^2-2 \theta \thetab)^{-(\D+\D_{ij})/2}   \Ocal_j(0)$, following the conventions of \cite{Penedones:2015aga}. In practice this gives us a very straightforward definition of $\OPE$:
	\be
	\left. (\partial^\a \partial_\a + a \ \partial^{\theta}\partial^{\thetab})(x^2-2 \theta \thetab)^{-\frac{\D+\D_{ij}}{2}} \right |_{\theta,\thetab=0}
	\equiv \OPE(\D_{ij}) \ (x^2)^{-\frac{\D+2+\D_{12}}{2}} \ .
	\ee
	The result of this computation is 
	\be
	\label{OPEcoeff}
	\OPE(\D_{ij})=\left(\Delta _{i j}-\Delta \right) \left(\Delta +\Delta _{i j}\right) \ .
	\ee
	By using formula \eqref{c20FromMultDecomp} and replacing \eqref{NORMcoeff} and \eqref{OPEcoeff},  we finally recover the exact form of $c_{2,0}(\ell=0)$, as predicted by \eqref{coefficientsCBdirred}.
	One could, of course, recover all the other coefficients (also for generic $\ell$) through similar computations. We did not invest time in doing so.

	\subsection{Comments}
	\label{sec:comments}
	In section \ref{CB_red} we showed that each superconformal block of the Parisi-Sourlas theory reduces to a single CB$_{d-2}$. This means that in the OPE all the superdescendant exchanges are reduced to descendant exchanges.  As we explained in sections \ref{dimredsec} and \ref{OPE_red}, this is somewhat non-trivial because contrary to how reductions work for trivial defects.
	In fact typically we would expect that a given supermultiplet would reduce to an infinite tower of CFT$_{d-2}$ multiplets. Our analysis shows that the infinite tower decouples inside the dimensionally reduced four point function, leaving only the contribution of a single multiplet.
	In the language of \eqref{OPEparallel1}, this is the consequence of the fact that the $S_1$ contributions drops out. It was already explained there but we would like to emphasize this again.
	We thereby obtain that the superconformal block decomposition of a scalar four point function reduces as follows:
	\be
	\label{red:CB_decomp}
	\la \Ocal_1(P_1) \dots \Ocal_4(P_4) \ra
	= \sum_{\D,\ell}  \l_{12 \, \D,\ell}\l_{\D,\ell \, 34} \, G^{d|2}_{\D,\ell}(P_i) 
	\longrightarrow
	\la \hat O_1(\hat P_1) \dots \hat O_4(\hat P_4) \ra  = \sum_{\D,\ell} \l_{12 \, \D,\ell}\l_{\D,\ell \, 34} \, G^{d-2}_{\D,\ell}(\hat P_i) \ ,
	\ee
	where $\hat P_i$ are the dimensionally reduced counterparts of $P_i$. The two sums run over the same spectrum, meaning that the exchanged operators are labelled by the same values $(\D,\ell)$. Furthermore all OPE coefficients match. 
	In addition, in section \ref{stress_tensor_red}, we detailed how the reduction of the super-stress tensor implies that the CFT$_{d-2}$ is local.
	These results explicitly show in which sense a  sector of the Parisi-Sourlas theory, given by $OSp(2|2)$-singlet operators, is described by a local non-supersymmetric CFT$_{d-2}$.
	A few more comments related to this picture are in order.
	
	From what we say above, and from the definition \eqref{reductionO}, it may seem that any superprimary operator in the SUSY theory descends to a primary operator in the CFT$_{d-2}$ with the same dimension $\D$ and which transform  in a representation of $SO(d-2)$ labelled by the same Young tableau. However this is not always true. Some superprimaries have no description in terms of primaries of the CFT$_{d-2}$. This happens when the Young tableau of $OSp(d|2)$ cannot be associated to a Young tableau of $SO(d-2)$.  Indeed Young tableaux of $SO(d-2)$ only have $[\frac{d}{2}-1]$ lines while the ones of $OSp(d|2)$ are of the form \eqref{OSpYT}. So, for example, in the Parisi-Sourlas theory there are operators labelled by a column of $n$ graded-antisymmetric boxes (for generic values of $n$), which are not there in the CFT$_{d-2}$ when $n$ is sufficiently large. Indeed one can see that the procedure \eqref{reductionO}, to generate a singlet of $OSp(2|2)$, would project such operators to zero.
	In other word, we cannot associate to all the supersymmetric operators an $OSp(2|2)$ singlet, hence the spectrum of the CFT$_{d-2}$ does not contain information of all the superprimaries of the Parisi-Sourlas theory.
	
	Let us discuss some simple consequences of this fact. 
	The first trivial comment is that, if a superprimary $\Ocal'$ is projected to zero by \eqref{reductionO}, then all correlation functions which contain $\Ocal'$ are projected to zero. One may then ask what happens when $ \Ocal'$ is exchanged inside the OPE of two superprimary operators $\Ocal_1,\Ocal_2$ which are not projected to zero. The answer is that the exchange of $ \Ocal'$ would be projected to zero. This is bound to happen since it is not possible to build the $SO(d-2)$ tensor structure which multiplies the projected operator $\hat {O'}$ inside the OPE.\footnote{
	E.g.~we can consider $\Ocal_1,\Ocal_2$ to transform in a spin $\ell$ representations in $5d$. In their $5$-dimensional OPE we may find operators $\Ocal'$ labeled by two equal rows of length $\ell$ in the Young tableau. While this representation exists in $5d$, it does not exist in $3d$ (for $\ell>1$). For $\ell=1$ instead this may be dualized to a parity odd vector representation, which does exist in $d=3$. See also footnote \ref{note:dual}.
	} 
{The associated partial wave vanishes since its leading OPE, which sets the boundary condition of the Casimir equation, is zero.}
	
	Finally it is interesting to wonder if the opposite of dimensional reduction may also work. Namely, given an abstract local CFT$_{d-2}$, can we always uplift it to a superspace $\mathbb{R}^{d|2}$ and obtain a Parisi-Sourlas CFT? This procedure indeed looks viable since  a huge part of the SUSY CFT data is described by the lower dimensional theory. 
	For example all four point functions of scalar operators of the CFT$_{d-2}$ can be easily uplifted preserving all CFT axioms. In fact it is clear that we can change the direction of the arrow in \eqref{red:CB_decomp}.
	However, the presence of the operators of the kind $\Ocal'$, discussed above, stops us from making the strong claim, since information about them cannot be obtained from lower dimensions. One could hope that this part of the spectrum can be fixed by imposing extra consistency conditions, as we speculate in the conclusions.
		Because of this complication, reconstruction of the full SUSY CFT$_d$ from CFT$_{d-2}$ is not a completely trivial problem and requires further analysis. This problem has an interesting analogy in the problem of critical dynamics, which we review in appendix \ref{sec:dyn}.

\section{Conclusions }\label{sec:conclusions}
In this paper we studied the implication of Parisi-Sourlas superconformal symmetry. 
This is conjectured to be realized in RG fixed points of models with random field type of disorder.\footnote{This should be distinguished from random bond type of disorder, which corresponds to disorder in the coupling $J$ in Hamiltonian \eqref{Ham}. Random-bond disorder is also interesting but the physics involved is quite different; see \cite{Komargodski:2016auf} for recent work.}
The full Parisi-Sourlas conjecture works in some cases and to fail in others, as we will revisit in a forthcoming paper. Here we narrowed the scope, and studied the problem from a different point of view, by focussing on the supersymmetric fixed points themselves.

These fixed points enjoy $OSp(d+1,1|2)$ superconformal symmetry.
They possess some unusual features which escape the conventional classifications of superconformal theories available in the high-energy physics literature.\footnote{E.g.~the classification of  \cite{Cordova:2016emh} assumes that the generators of the supersymmetries transform in the fundamental spinor representation of the Lorentz group.
 One may be confused since the algebra $OSp(d+1,1|2)$ also describes some unitary theories, as the six-dimensional $(2, 0)$ SCFTs. However, even when the algebra is of the Parisi-Sourlas type, the generators of the superconformal symmetries classified by \cite{Cordova:2016emh} are embedded inside the orthosymplectic rotations in a different way, see e.g. \cite{Arvidsson:2006tk}.
}  In fact Parisi-Sourlas theories are non-unitary and the generators of their supersymmetries transform not as spinors but as scalars or vectors under rotations. While lack of unitarity may seem unusual to a particle physicist, unitarity is not a request of fundamental importance in statistical physics context, and numerous non-unitary theories are known to play a role in nature.

In section \ref{sec:susy} we showed how these superconformal theories work from an axiomatic point of view.
We started with a detailed analysis of their symmetries and representations. We explained how to embed the stress tensor of the theory in a superconformal multiplet. 
We then analyzed correlation functions by introducing a super-embedding space formalism and we exemplified it in some simple cases. Finally we showed how to define the OPE and the superconformal block decomposition of four point functions. While this analysis could have revealed some pathological features, it did not: these theories seem perfectly healthy. 

The most compelling aspect of Parisi-Sourlas supersymmetric theories is that they undergo the so called ``dimensional reduction''. Namely that a sector (the $OSp(2|2)$-singlet sector) of the SUSY theory is described by a  $d-2$ dimensional CFT. 
In section \ref{sec:dimred} we carefully explain the meaning of dimensional reduction and its implications at the level of axiomatic CFTs.
We showed that the locality of the Parisi-Sourlas CFT implies the locality of the $d-2$ dimensional theory.
We then explained that every superprimary descends under dimensional reduction to one primary, while infinitely many additional operators which would naively also be expected to arise (as they do arise for reductions to trivial defects) decouple. 
This is explained by direct inspection of the OPE and by the computation of the superconformal blocks of the theory, which we prove to be equal to the usual conformal blocks in $d-2$ dimensions. 
Finally, from the study of superconformal blocks, we obtained a new finite-term linear relation involving conformal blocks in $d$ and $d-2$ dimensions. This relation is between non-supersymmetric blocks and may look like it has nothing to do with supersymmetry. However, the existence of this surprising relation is a direct consequence of the Parisi-Sourlas supersymmetry and could not have been guessed otherwise.

We have thus illustrated that the Parisi-Sourlas CFTs are axiomatically well defined and that they undergo dimensional reduction. These facts have some important consequences. 
First we confirm that such SUSY fixed points do exist, hence they could be reached at the end of an RG flow. One reason why they are not always reached at the end of the random field theory flow could be their RG instability. If this is the case, it follows that they should be reachable by tuning extra parameters (e.g. by changing the disorder distribution). We find this hypothesis very tempting and it would be really exciting to see experimental and numerical evidences of this fact. We will have more to say in \cite{paper2}.

A straightforward extension of this work is to study theories with superconformal symmetry $OSp(d-1+2n,1 | 2n)$. Loosely speaking, here the $2n$ fermionic  degrees of freedom should effectively cancel with $2n$ bosonic ones. So with an analysis analogous to ours it should be possible to see that these theories undergo a dimensional reduction by $2n$. Perhaps such theories also describe observables of some critical disordered theories.

 We would like to mention here another, seemingly unrelated, known class of examples, when supersymmetric theories reduced to a trivial defect also give a local theory: $\Ncal=2$ superconformal theories in $d=4$ and six-dimensional $(2,0)$ theories in $d=6$. Reduced to a plane, they give 2d chiral algebras possessing a local stress tensor \cite{Beem:2013sza,Beem:2014kka}. These chiral algebras capture correlation functions of certain supersymmetrically protected operators put in the plane and contracted with $x$-dependent polarization vectors (``twisted-translated''). In \cite{Beem:2013sza,Beem:2014kka}, the higher dimensional SCFT is unitary, while the reduced chiral algebra is non-unitary.
This is opposite to our case, when the higher dimensional theory is always non-unitary, while the reduced theory may well be unitary.

Finally we want to comment on a very deep but speculative direction hinted at in our work.
From our analysis there seems to exist a very general relation which connects CFTs in different dimensions. 
Indeed we found that the dimensionally reduced CFT$_{d-2}$ captures a huge part of the CFT data of the Parisi-Sourlas theories. For example the whole spectrum of scalar superprimaries and all the three point function coefficients appearing in their OPEs is determined by the CFT$_{d-2}$. 
This seems to suggest that the opposite of dimensional reduction (``dimensional lift'') may also work. Namely that any given  CFT$_{d-2}$ can be lifted to $\mathbb{R}^{d|2}$ to define a theory with $OSp(d+1,1 | 2)$ supersymmetry.
(This idea can also be inspired by an analogy in the problem of critical dynamics reviewed in appendix \ref{sec:dyn}.)
However, as we explained in section \ref{sec:comments}, this uplift is not trivial because some superprimary operators of the Parisi-Sourlas theory have no counterpart in the CFT$_{d-2}$.
It would be interesting to see if this part of the spectrum can be reconstructed from crossing symmetry consistency conditions. For example, it should be possible to constrain the CFT data of these operators by bootstrapping their correlation functions in the higher dimensional theory.
If this picture is proven correct we would conclude that the existence of a local CFT$_{d}$, automatically implies the existence of a discrete sequence of CFT$_{d+2n}$ for integer $n$, which are local, non-unitary and supersymmetric.
 Alternatively, if the conjecture is incorrect, there should be a condition which determines which CFTs can be lifted and which cannot. Either way this is exciting and deserves further investigation.

\section*{Acknowledgment}
S.R.~is grateful to \'Edouard Br\'ezin for the introduction to Random Field Ising Model, to Giorgio Parisi for the invitation to the stimulating workshop \emph{Beyond Mean Field Theory} (Rome, January 2018), and to Nicolas Sourlas for many stimulating conversations, for mentioning \cite{CARDY1983470}, and for comments on the draft. A.K.~and E.T.~would like to thank R. Gopakumar, N. Ishtiaque, B. Le Floch, G. Mandal, M. Meineri, M. Paulos, J. Penedones, B. van Rees, V. Schomerus, D. Simmons-Duffin, A. Sinha and participants of  \emph{Bootstrap 2019} (Perimeter Institute) for useful discussions. A.K. would like to thank YITP, Stony Brook for hospitality during the course of this work.
E.T. would like to thank the participants of the workshop \emph{Nonperturbative Methods for Conformal Theories} (Natal, April 2019). This work was partially supported by the Simons Foundation grant 488655 (Simons Collaboration on the Nonperturbative Bootstrap) and by Mitsubishi Heavy Industries (MHI-ENS Chair).

	\appendix
	
\section{Perturbative dimensional reduction}
\label{app:pertdimred}
The equivalence of the supersymmetric theory in \eqref{Ssusy} and the $d-2$ dimensional theory \eqref{Sdm2} can be shown from perturbation theory. 
In this appendix we explain how to recover \eqref{relcorr} by showing that any general Feynman integral from the SUSY action reduces to a Feynman integral of the $(d-2)$-theory. 

We assume a general polynomial form of $V(\Phi)=\sum_{m\ge 2}g_m \Phi^m$\,. Then a generic superfield diagram would be:
\be\label{FeynGen}
F(y_1,\cdots,y_n)=(2\pi)^{p_{max}}\int [\dd y_p]\prod_{i<j}\big(G_{\Phi\Phi}(y_{ij})\big)^{q_{ij}}\,.
\ee
Here $[\dd y_p]=\prod_p \dd^dx_p \dd\thetab_p\dd \theta_p$, and $p \in {\bf I} =\{1, \dots p_{max}\}$ labels the internal points (vertices) of the diagrams.
We shall use indices $i,j,\ldots$ to number generic points, internal or external. Also let us call the total number of propagators as $N$. The factor outside the integral arises from rescaling of the coupling constants due to the overall factor in the SUSY action \eqref{Ssusy}.  The number $q_{ij}$ denotes the power with which a certain $G_{\Phi\Phi}(y_{ij})$ arises, where $y_{ij}=y_i-y_j$ and
$G_{\Phi\Phi}$ is the free theory propagator of the superfield $\Phi$. We can obtain $G_{\Phi\Phi}$ (with the appropriate normalization factor) from \eqref{Ssusy} by setting $V(\Phi)=0$\,,
\begin{align}
G_{\Phi\Phi}(y)
=\frac{a}{\big(x^2+2\thetab\theta \big)^{\frac{d-4}{2}}}\,,
		\end{align}
where $a\equiv \Gamma \left(\frac{d}{2}-2\right)/(8\pi ^{\frac{d}{2}} )$. For simplicity we are using massless propagators and treating mass as a perturbation.  In what follows it will be convenient to split the coordinates in different groups:
\begin{align}\label{brkdn}
y^a \equiv \{x^\a,\theta,\thetab\}=\{{\hat x}^\a,{x^\perp}^\a,\theta,\thetab\}\,,
\qquad \ \
\hat x^\a \equiv \{x^1,\cdots,x^{d-2}\} \ ,
\qquad \ \
 {x^\perp}^\a \equiv \{x^{d-1},x^{d}\}\,.
\end{align}
In order to prove dimensional reduction of \eqref{FeynGen}, we want to show that if we set ${x_i^{\perp}}^\a=\theta_i=\thetab_i=0$ for any external point $y_i$, the integral reduces to
\be\label{toprove}
F(\hat x_1,\cdots,\hat x_n)=\left( 2\pi \right)^{p_{max}}\int [\dd \hat x_p]\prod_{i<j}\big(G_{\hat \phi\hat \phi}(\hat x_{ij})\big)^{q_{ij}}\,,
\ee
where $[\dd \hat x_p]=\prod_p \dd^{d-2} \hat x_p$. The function  $G_{\hat \phi\hat \phi}(\hat x_{12})=a \, {(\hat x_{12}^2)}^{-\frac{d-4}{2}}$ is the free theory propagator of $\hat \phi$ from \eqref{Sdm2}\,. Notice that the normalizations of the propagators of $\hat \phi$ and $\Phi$ are the same, which is crucial to the simplification of the computations. This follows from Fourier transforming their respective momentum space propagators as a direct consequence of \eqref{Radial2}.\footnote{It is possible to normalize both propagators to one, which amounts to rescaling the kinetic terms in the two actions \eqref{Ssusy} and \eqref{Sdm2}.}

A useful tool to handle propagators in position space is the Schwinger  parametrization: 
\be
\frac{1}{A^r}=\frac{1}{\G(r)}\int_0^\infty du \ u^{r-1}e^{-u A}\,.
\ee
Introducing one Schwinger parameter $u_{ij}$ per propagator, Eq.~\eqref{FeynGen} becomes
\begin{align}\label{Fgen2}
F(y_1,\cdots,y_n)=\Ncal  a^{N}(2\pi)^{p_{max}} \int [\dd y_p] \int [\dd u_{ij}] \prod_{i<j}\Big(u_{ij}^{q_{ij}(\frac{d}{2}-3)}\Big) \exp\bigg[-{\sum_{i,j}\big(x_i \cdot x_j M_{ij}+2\thetab_i \theta_j M_{ij}\big)}\bigg]\,.
\end{align}
Here $[\dd u_{ij}]=\prod_{i<j}\dd u_{ij}$, $\Ncal$ is a product of Gamma functions, and  the matrix $M_{ij}$ is given by (this matrix is obtained by expanding $x_{ij}^2$ and $\theta_{ij}\bar\theta_{ij}$ in terms of $x_i$, $\theta_i$, $\bar\theta_j$):
\be
M_{ij}=\begin{cases}
	\sum_k u_{ik},& \text{if } i=j\\
	-u_{ij} \ (\equiv -u_{ji}),              & \text{otherwise}\,.
\end{cases}
\ee
Now separating the variables as in \eqref{brkdn} we can rewrite \eqref{Fgen2} as
\begin{align}
\begin{split}
F(y_1,\cdots,y_n)&=\Ncal a^{N}(2\pi)^{p_{max}}\int [\dd u_{ij}] \prod_{i<j}\Big(u_{ij}^{q_{ij}(\frac{d}{2}-3)}\Big) \int [\dd \hat x_p]   \exp\bigg[-{\sum_{i,j}\hat x_i \cdot \hat x_j M_{ij}}\bigg]\\
&\times \int [\dd x^{\perp}_p][\dd \bar{\theta}_p][\dd \theta_p]  \exp\bigg[-{\sum_{i,j}\big(x^{\perp}_i \cdot x^{\perp}_j M_{ij}+2\thetab_i\theta_j M_{ij}\big)}\bigg]\,.
\end{split}
\label{divided_integral}
\end{align}
Let us focus on the integral in the second line of \eqref{divided_integral}. 
By setting ${x_i^\perp}^{\a}=\theta_i=\thetab_i=0$ for the external points, only internal points contribute to the sum over $i,j$.
Namely the integral becomes
\be
\mathcal{I}= \int [\dd x^{\perp}_p][\dd \bar{\theta}_p][\dd \theta_p]  \exp\bigg[-\sum_{p_1,p_2\in {\bf I}}\big(x^{\perp}_{p_1} \cdot x^{\perp}_{p_2} \tilde{M}_{p_1 p_2}+2\thetab_{p_1}\theta_{p_2} \tilde{M}_{p_1 p_2}\big)\bigg]\, ,
\ee
where ${\bf I}$ is the set of internal points and $\tilde{M}_{p_1 p_2}$ is a $p_{max} \times p_{max}$ matrix (recall that $p_{max}$ is the number of internal points) obtained by restricting $M_{ij\in\bf I}$. Now, since the integrals over $x^\perp, \theta_a, \bar{\theta}_a$ are Gaussian, $\mathcal{I}$ can be computed as follows:
\be
\mathcal{I}= \left[\frac{ \pi^{p_{max}}}{\det (\tilde M)} \right] \times \left[\frac{\det (\tilde M) }{\pi^{p_{max}} }\right] =1 \, .
\ee
Here the results in the left (right) square brackets is obtained from the bosonic (fermionic) integral. 
We therefore conclude that 
\begin{align}
F(\hat x_1,\cdots,\hat x_n)&=\Ncal a^N\left( 2\pi \right)^{p_{max}}\int [\dd u_{ij}] \prod_{i<j}\Big(u_{ij}^{q_{ij}(\frac{d}{2}-3)}\Big) \int [\dd \hat x_p]   \exp\bigg[-{\sum_{i,j}\hat x_i\cdot\hat x_j M_{ij}}\bigg]\nonumber\\
&=\left( 2\pi\right)^{p_{max}} \int [\dd \hat x_p]\prod_{i<j}\big(G_{\hat \phi\hat \phi}(\hat x_{ij})\big)^{q_{ij}}\,.
\end{align}
So  we get the same Feynman integral as in $d-2$ dimensions. Although it may not be obvious, the mechanism for dimensional reduction is similar to what was demonstrated in \eqref{RadialInt1}. Namely, for integrals of functions of distances, the fermionic components cancel out with two of the bosonic ones, which results in dimensional reduction.

\section{Useful relations in super(embedding)space}
\label{app: superspace}
In order to make more clear of the computations of section \ref{sec:dimred} we collect here some useful properties of the superspace and the super-embedding space.

We shall first consider  the super-embedding space $\mathbb{R}^{d+1,1|1}$.
The metric $\bar g_{AB}$ defined in \eqref{s_embedding_metric} is non-diagonal and non-symmetric because of the $J_2$ piece so we need to be careful with the position of the indices:
\be
\label{commutation super-embedding}
P^A P^B=(-1)^{[A][B]}  P^B P^A  \ ,
\qquad 
\bar g_{AB}=(-1)^{[A][B]} \bar g_{BA} \ ,
\ee
where $[A]=0$ if $A = 0, \dots ,d+1$ and $[A]=1$ for the remaining two grassmannian coordinates.
Equations \eqref{commutation super-embedding} in turn imply that the scalar product is symmetric $P\cdot Q=Q \cdot P$.
In order to raise and lower  indices we use the following conventions:
\be
P_A \equiv \bar g_{A B} P^B  \ , \qquad P^A \equiv \bar g^{B A} P_B  \ .
\ee
By using the above formulas on the metric itself we obtain that $\bar g^A_{\phantom{A}B}=\d^A_B$ and $\bar g_B^{\phantom{B}A}=\d_B^A (-1)^{[A][B]} $. Hence $\bar g^A_{\phantom{A}A}=d+4$ while $\bar g_B^{\phantom{B}B}=d$.
Furthermore it is important to remember that $\bar g^{AB}$ is not the inverse of $\bar g_{AB}$. In fact 
\be
\bar g^{AB} \bar g_{B C}=(-1)^{[B] ([A]+[C])} \bar g^{BA} \bar g_{CB} =(-1)^{[B] (2 [A])} \bar g^{BA} \bar g_{CB}=\bar g_{C}^{\phantom{C} A} = (-1)^{[A] [C]}  \delta^{A}_{\phantom{A} C}
\ .
\ee
Let us now exemplify how the derivatives
$
\partial_{M}\equiv\partial_{P^M}=(\partial_{P^0},\partial_{P^\m},\partial_{P^{d+1}},\partial_{\theta} \, ,\partial_{\bar \theta}) 
$
act in super-embedding space. It is again important to keep track of the order of the indices in all computations.
For example,
\be
\partial_{M} P^{N}= \d_{M}^N \ , \quad
\partial_{M} P_{N}= \bar g_{N M} \ , \quad
\partial^{M} P^{N}= \bar g^{N M} \ , \quad
\partial^{M} P_ {N}= (-1)^{[M][N]} \d^{M}_ N  \ .
\ee
All the previous equations can be alternatively defined in superspace $\mathbb{R}^{d|2}$. The results are basically the same after the map:
\be
\bar g^{AB} \to g^{ab} \, , 
\qquad
P^A \to y^a\, ,
\ee 
where $y^a=(x^\alpha,\theta,\bar\theta)\in\mathbb{R}^{d|2}$ are the superspace points and $g^{ab}$ is the superspace metric defined in \ref{metricOSp}.
In particular let us emphasize that by taking different traces of the superspace metric the following two results can be obtained:
\be
g^a_{\phantom{a}a}=d+2 \, ,
\qquad
g_a^{\phantom{a}a}=d-2\,.
\ee
Finally we stress that the trace $g^a_{\phantom{a}a}$ depends on the choice of the tensor basis. The appropriate $OSp(d|2)$ invariant trace, the so called \emph{supertrace}, is $g_a^{\phantom{a}a}$. It is only the supertrace that can appear in our computations.  

\section{Example: Free theory}
\label{Free_ex}
As a demonstration let us consider the supersymmetric theory of a free massless scalar $\Phi$, which can be written as
\be\label{freeth}
S_{SUSY}=2\pi\int d^dx d\thetab d\theta  \ \frac{1}{2} g^{ab}\partial_a\Phi\partial_b\Phi \,.
\ee
Because of the $\theta,\thetab$ integrals, the dimension of the super-Lagrangian is $d-2$, which implies that the dimension of  $\Phi$ is  $\frac{d}{2}-2$. 
Indeed this is the dimension of a scalar superprimary that satisfies the equations of motions $\partial^a \partial_a \Phi=0$, as we explained in subsection \ref{subsec:symCFT}.
By expanding $\Phi$ in $\theta$ and $\thetab$ (see equation \eqref{def:Phi}), we conclude that the dimensions of its constituents are
\be
\big[\varphi\big]=\frac{d}{2}-2 \, ,
\qquad
\big[\psi\big]=\big[\,\psib\,\big]=\frac{d}{2}-1 \, ,
\qquad
\big[\omega\big]=\frac{d}{2} \, .
\ee

Noether's theorem can be used to define a super-stress tensor:
\be\label{stressform}
{\mathcal{T}}^0_{ab}=\frac{\d \mathcal{L}}{\d(\partial^a \Phi) }\partial_b \Phi-g_{ba}\mathcal{L}\, ,
\ee
with the conservation law, $\partial^a{\mathcal{T}}^0_{ab}=0$.
From \eqref{stressform} it is clear that $[{\mathcal{T}}^0]=d-2$, as predicted in  section \ref{subsec:symCFT}. 
We construct an improved version of the super-stress tensor that satisfies the tracelessness property ${\mathcal{T}^a}_a=0$:
\be
\mathcal{T}_{ab}=(\partial_{(a}\Phi)(\partial_{b]}\Phi)- \frac{d-4}{d-2}\Phi\partial_{a}\partial_{b}\Phi-\frac{g_{ba}}{d-2}(\partial^c \Phi) (\partial_c \Phi) \, ,
\ee
where $(\cdot]$ implements graded-symmetrization.
It is easy to verify, following the equation of motion of \eqref{freeth}, that this satisfies a super-conservation equation like \eqref{supcons}\,. Expanding in $\theta$, $\bar{\theta}$, we get
\begin{align}
{\mathcal{T}^{\mu\nu}_{\theta\bar{\theta}}}=T^{\mu\nu}=&2 \partial^{(\mu} \varphi \partial^{\nu)} \omega +2 \partial^{(\mu} {\psi}\partial^{\nu)} \psib+\frac{4-d}{d-2}(\varphi\partial^\mu\partial^\nu\omega+\omega\partial^\mu\partial^\nu\varphi+\psi\partial^\mu\partial^\nu \psib-\psib\partial^\mu\partial^\nu \psi ) \nonumber \\ &-\frac{2g^{\mu\nu}}{d-2}(\partial_{\rho}\varphi\partial^{\rho}\omega+\partial_\rho {\psi} \partial^\rho \psib)\,.
\end{align}
This satisfies the conservation equation $\partial_\mu T^{\mu\nu}=0$ and hence is the stress-energy tensor of the $d$-dimensional theory that one obtains from \eqref{freeth} by performing the $\theta$, $\bar\theta$ integral\,.

Now let us use the formula \eqref{lowerstress} to obtain the $SO(d-2)$ stress-tensor. This is given by
\be
\hat T_{\a\b}=({\partial}_{\a}\varphi)({\partial}_{\b}\varphi)- \frac{d-4}{d-2}\varphi{\partial}_{\a}{\partial}_{\b}\varphi-\frac{g_{\a\b}}{d-2}({\partial}_\g \varphi) ({\partial}^\g \varphi) \,.
\ee
Here $\a$ runs from $1$ to $d-2$. The above satisfies the conservation equation $\partial^\a \hat T_{\a\b}=0$ in $d-2$, given the theory is free, i.e. $\partial^2 \varphi=0$\,.
\vspace{1cm}
	
\section{Supersymmetry in the problem of critical dynamics}	\label{sec:dyn}

In this paper we dealt with the appearance of Parisi-Sourlas supersymmetry in physics of disordered systems.
In this appendix we will review how supersymmetry plays a role in another problem of statistical physics: critical dynamics. Our review is based on \cite{ZinnJustin:2002ru}, chapters 36, 16, and 17. See also \cite{cardy_1996}, chapter 10.

Consider a $d$-dimensional system at a thermodynamic continuous phase transition. We are interested in correlation functions of local operators, e.g. the fluctuating order parameter field $\phi$. The usual problem of critical statics concerns equilibrium correlation functions, determined by the equilibrium Gibbs distribution. E.g.~for the ferromagnetic phase transition we might consider a $d$-dimensional path integral with the Landau-Ginzburg action $S[\phi]=\int d^dx\,H[\phi]$, $H(\phi)=(\partial \phi)^2 +m^2 \phi^2+\lambda \phi^4$, tuning $m^2$ to reach the IR fixed point described by a CFT$_{d}$. Time dependence is, by definition, absent in critical statics. Physically, we are dealing with equal-time correlators: 
\be
\langle \phi(x_1) \ldots \phi(x_n)\rangle_{\rm CFT} = \langle \phi(x_1,t) \ldots \phi(x_n,t)\rangle\,,
\label{equal}
\ee
where the average is over equilibrium state, which is time independent.

The problem of critical dynamics is more general and concerns correlators at unequal times, i.e.
\be
\langle \phi(x_1,t_1) \ldots \phi(x_n,t_n)\rangle\,.
\label{unequal}
\ee
When we insert the operator $\phi(x_1,t_1)$ at the smallest time $t_1$ (say), this disturbs the system away from equilibrium and it starts relaxing back to it, so that when we complete the measurement by inserting other operators at later times we probe a perturbed system.
If all operators are inserted at the same time $t_1=\ldots=t_n$ this perturbation has not spread and we measure the static correlators as before, but for unequal times we will measure different quantities which will depend on time differences. 

Unequal-time correlators \eqref{unequal} are expected to be scale invariant under the transformation
\be
x\to\lambda x,\quad t\to \lambda^z t\quad(\lambda>0),
\ee
where $z$ is a parameter called the dynamical critical exponent. Unlike for static correlators \eqref{equal}, there is no reason to expect any more complicated extended symmetry such as conformal invariance.

While equilibrium equal-time correlators \eqref{equal} depend only on the Gibbs distribution, unequal-time correlators \eqref{unequal} depend on the additional piece of data: the mechanism by which the system relaxes back to equilibrium. Various such mechanisms are listed in the literature, but two most well-known ones are Model A and Model B. Model B conserves the average value of the order parameter, while Model A (also known as the Glauber dynamics) does not. When doing Monte Carlo simulation of the lattice Ising model, Model A is the usual Metropolis algorithm flipping individual spins, while Model B is flipping only opposite sign pairs of nearby spins: $+-\leftrightarrow -+$.
Universality holds for critical dynamics, with models differing by symmetry considerations, like Model A and Model B, corresponding to different universality classes. Thus each static universality class may correspond to several dynamic universality classes, which will have different unequal-time correlators \eqref{unequal}, but will share the same equal-time correlators \eqref{equal}. In particular, they have a different dynamical critical exponent $z$.

In the continuum description, approach to equilibrium can be described by the Langevin equation
\be
\partial_t \phi(x,t) = - \frac{\delta}{\delta \phi(x,t)} H[\phi] +\nu(x,t)\,,
\label{Langevin}
\ee
where $\nu$ is a Gaussian white noise, $\langle \nu(x,t)\nu(x',t')\rangle = 2\delta(x-x')\delta(t-t')$ (for this normalization of its two-point function, the equilibrium Gibbs distribution is precisely $e^{-S[\phi]}$). This equation does not preserve the average $\phi(x,t)$ so it corresponds to Model A. For other models the story would be similar with a somewhat different Langevin equation, see \cite{ZinnJustin:2002ru}, \cite{cardy_1996}.

The next step is to encode \eqref{Langevin} in a path integral. For this we introduce a Lagrange multiplier field $\omega$ and two Grassmann fields $\bar c$ and $c$ to reproduce the determinant arising from the functional $\delta$-function. Integrating over the noise, we get a path integral with the action in $d+1$ dimensions (see \cite{ZinnJustin:2002ru}, (16.128))
\be
\mathcal{S}[\phi] =\int d^dx dt \{-\omega ^2 +\omega [\dot \phi +  \delta H/\delta \phi ]-c (\partial_t
+\delta^2 H/\delta \phi \delta\phi) \bar c\}\,.
\ee
One then introduces the superfield
\be
\Phi(x,t,\theta,\bar\theta)=\phi(x,t)+\bar \theta c(x,t)+ \theta \bar c(x,t) + \theta\bar\theta \omega(x,t)
\ee
and rewrites the action as
\be
\mathcal{S}[\phi]=\int d^dx\, dt\, d\bar\theta\, d\theta \{ \bar D\Phi D\Phi +H[\Phi]\}\,,
\ee
where the superderivatives are $\bar D=\partial_{\bar\theta}$, $D=\partial_{\theta}-\bar\theta \partial_t$. This action has two supersymmetry generators $Q=\partial_{\theta}$, $\bar Q=\partial_{\bar\theta}+\theta \partial_t$.

This supersymmetry has a physical raison d'\^etre. Any problem of critical dynamics must satisfy a physical constraint known as the fluctuation-dissipation theorem, which expresses the two point function $\langle \phi(x_1,t_1)\phi(x_2,t_2)\rangle$ as an integral of a response function (see \cite{cardy_1996}). In the supersymmetric formulation, the response function is the two point function $\langle \phi(x_1,t_1)\omega(x_2,t_2)\rangle$, and the fluctuation-dissipation theorem can be recovered as a supersymmetric Ward identity.

The problem of critical dynamics thus consists in taking a $d$-dimensional CFT describing critical statics and in finding a $(d+1)$-dimensional, space+time, supersymmetric theory which reduces back to the CFT when all times are set equal. Stated this way, this problem is analogous to the problem of ``dimensional lift'' which we encountered in the random field Ising model context in section \ref{sec:comments} and in the conclusions. We find this analogy quite suggestive, even though it is not complete, most notably because of the absence of any symmetry mixing $x$ and $t$ in the critical dynamics case. 

It should be noted that the $d\to d+1$ connection does not play much of a role in the currently existing practical methods of solving critical dynamics, which basically analyze the $(d+1)$-dimensional RG problem from scratch. It would be interesting to find an alternative method which would take into account the $d$-dimensional static information when it is available (e.g.~when critical statics is exactly solved, as for the 2d Ising model).

	\small
	\parskip=-10pt
	\bibliography{references}
	\bibliographystyle{utphys}

\end{document}